\documentclass{aa}  
\usepackage{graphicx}
\usepackage{xcolor}
\usepackage{txfonts}
\usepackage{placeins}
\usepackage{lscape}
\usepackage{float}
\usepackage{multicol}

\begin{document} 

\title{New kinematic map of the Milky Way bulge
\thanks{Based on observations taken at the ESO Very Large Telescope with the MUSE instrument under program IDs 101.B-0381(A) (SM; PI: Zoccali), and 101.D-0363(A) (SM; PI: Minniti).}}

   \author{C. Quezada\inst{1,2}
          \and
          M. Zoccali\inst{1,2}
          \and
          E. Valenti\inst{3,4}
          \and
          A. Rojas Arriagada\inst{2,5,6,7}
          \and
          A. Renzini\inst{8}
          \and
          O.A. Gonzalez\inst{9}
          \and
          A. Mucciarelli\inst{10, 11}
          \and
          M. Rejkuba\inst{3}
          \and
          F. Surot\inst{12}
          \and
          A. Valenzuela Navarro\inst{1, 2}
          }

\institute{Instituto de Astrof\'isica, Pontificia Universidad Cat\'olica de Chile, Av. Vicu\~na Mackenna 4860, 782-0436 Macul, Santiago, Chile
\email{ciquezada@uc.cl}
    \and
Millennium Institute of Astrophysics, Av. Vicu\~na Mackenna 4860, 82-0436 Macul, Santiago, Chile 
    \and
European Southern Observatory, Karl Schwarzschild-Strabe 2, 85748 Garching bei München, Germany
    \and
Excellence Cluster ORIGINS, Boltzmann\---Stra\ss e 2, D\---85748 Garching bei M\"{u}nchen, Germany
    \and
Departamento de F\'isica, Universidad de Santiago de Chile, Av. Victor Jara 3659, Santiago, Chile
    \and
N\'ucleo Milenio ERIS
    \and
Center for Interdisciplinary Research in Astrophysics and Space Exploration (CIRAS), Universidad de 
Santiago de Chile, Santiago, Chile
    \and
INAF – Osservatorio Astronomico di Padova, Vicolo dell’Osservatorio 5, I-35122 Padova, Italy
    \and
UK Astronomy Technology Centre, Royal Observatory, Blackford Hill, Edinburgh, EH9 3HJ, UK
    \and
Dipartimento di Fisica \& Astronomia "Augusto Righi" - Universitá degli Studi di Bologna, Via Piero Gobetti 93/2, 40129 Bologna, Italy
    \and
INAF - Osservatorio di Astrofisica e Scienza dello Spazio di Bologna, Via Gobetti 93/3, 40129 Bologna, Italy
    \and
Independent Astronomer
}

% \date{19th August 2025}

\abstract
{The kinematics of the Milky Way bulge is known to be complex, reflecting the presence of multiple stellar components with distinct chemical and spatial properties. 
In particular, the bulge hosts a bar structure exhibiting cylindrical rotation, and a central velocity dispersion peak extending vertically along the Galactic latitude.
However, due to severe extinction and crowding, observational constraints near the Galactic plane are sparse, underscoring the need for additional data to improve the completeness and accuracy of existing kinematic maps, and enabling robust comparison with dynamical models.
}  
{This work aimed to refine the existing analytical models of the Galactic bulge kinematics by improving constraints in the innermost regions. 
We present updated maps of the mean velocity and velocity dispersion by incorporating new data near the Galactic plane. 
}
{We combined radial velocity measurements from the GIBS and APOGEE surveys with both previously published and newly acquired MUSE observations. 
A custom\---developed Python\---based tool, {\tt PHOTfun}, was used to extract spectra from MUSE datacubes using PSF photometry based on DAOPHOT-II, with an integrated GUI for usability. The method included a dedicated extension, {\tt PHOTcube}, optimized for IFU datacubes. We applied Markov Chain Monte Carlo techniques to identify and correct for foreground contamination and to derive new analytical fits for the velocity and velocity dispersion distributions. 
Our analysis included nine new MUSE fields located close to the Galactic plane, bringing the total number of mapped fields to 57 including ~23000 individual RV measured.}
{The updated kinematic maps confirm the cylindrical rotation of the bulge and reveal a more boxy morphology in the velocity dispersion distribution, while preserving a well\---defined central peak. 
The {\tt PHOTfun} software, designed for flexible PSF photometry and spectral extraction from IFU data, is publicly available via pip for the community.}
{}


\maketitle

\section{Introduction}

One of the common massive components in galaxies is the bulge, typically defined as the 
central region where a substantial fraction of the stellar mass is concentrated, resulting 
in a high surface brightness. Moreover, bulges generally form alongside their host galaxies, 
retaining traces of their complete evolutionary history.

In the bulge, the spatial distribution of stars is a consequence of the kinematical evolution
during the galaxy formation history. 
An example is the so-called classical or spheroidal bulges seen in some spiral galaxies, which 
are pressure-supported structures traditionally believed to have formed through early, violent mergers 
\citep{kauffmann+93, hopkins+10, garrison-kimmel+18}. 
Another example is the elongated bar often seen crossing the center of spiral galaxies.

This feature is believed to arise from the secular evolution of the galactic disk, 
in which kinematic instabilities in stellar orbits give rise to a barred structure. 
Subsequent buckling instabilities may occur, resulting in a boxy or peanut-
shaped bulge \citep{patsis+02, athanassoula05, shen+2010, portail+2015}. 

Besides mergers and orbital instabilities, in more recent years a third option has gained 
momentum, one in which the formation of bulges in high-redshift galaxies can result from 
the inward migration of gas from the disk, then feeding a star-forming bulge. In one option 
of this scenario, giant star-forming clumps can form in the gas-rich disk, which would then 
migrate and coalesce toward the center 
\citep[for example, ][]{immeli+2004, carollo+07, elmegreen+08, genzel+2008, bournaud+2009}
Besides clump formation and migration, the possibility of overall violent disk instabilities 
has also been proposed, which would lead to the central accumulation of large amounts of gas, 
then rapidly depleted by star formation and winds 
\citep[for example, ][]{dekel+14, tacchella+16} Relatively smooth radial gas flows have 
been detected in high-redshift galaxies \citep{genzel+23}, possibly driven by bars or 
spiral arms, then contributing to bulge growth via in situ star formation.
In all such versions, rotating bulges form rapidly being fed by the disk, in a gas-rich, 
highly dissipative environment, for which ALMA spatially resolved observations have shown 
plausible examples among redshift $\sim 2$ galaxies \citep[for example, ][]{tadaki2017}.

In the Milky Way (MW), the bulge is the region within a radius of $\approx$3.5\,kpc around the center.
This central component concentrates a substantial fraction, $\sim$30\%, of the total stellar mass
of the Galaxy \citep{valenti+2016, portail+17barps, simion+2017, zoccali&valenti24}. 
It is one of the oldest galactic components (for instance, $\approx$10\,Gyrs), 
hosting some of the first stars formed in the Galaxy \citep{tumlinson+10, 
clarkson+11, bensby+2017, renzini+2018}. The bulge lies only 
about 5–8\,kpc from us, making it exceptionally well resolved 
in stars as faint as the main sequence; hence, it offers us a unique opportunity to study this
crucial component for understanding the MW formation

The elongated stellar orbits characteristic of a bar, coupled with a spheroidal spatial distribution,
constitute distinct kinematic and structural signatures of the bulge formation process, as reflected
in the line of sight velocity distributions. 
Consequently, constructing detailed maps of radial velocity (RV) and 
RV dispersion ($\sigma_{RV}$) serves as a
powerful diagnostic tool for constraining the formation history of the Galactic bulge. 
\citep{zoccali+14, molaeinezhad+16, simion+2017, debattista+2017, debattista+19}. 
Several spectroscopic surveys that have mapped the chemistry and kinematics of the MW bulge 
(for example, BRAVA: \citet{Rich+07BRAVA, kunder+2012}, ARGOS: \citet{freeman+13_argosII}, Gaia ESO:  
\citet{rojas-arriagada+2014}, GIBS: \citet{zoccali+14}, APOGEE: \citet{rojas-arriagada+2020}) 
have revealed that it hosts at least two stellar populations with distinct spatial 
distribution, chemical composition, and kinematics.
Specifically, while metal-rich stars trace the barred, boxy/peanut structure, the metal-poor 
population follows a more spheroidal distribution \citep{babusiaux+10, zoccali+2017, 
queiroz+20, lim+21}. 
Mapping the global kinematics of the MW bulge offers a well resolved benchmark for comparison
with more distant galaxies, where individual stars cannot be distinguished. Moreover, the presence 
of multiple structural components provides valuable insights into more intricate galaxy 
formation pathways.

Using measurements from the GIBS survey, \citet[][hereafter Z14]{zoccali+2014} presented the
first analytical maps of RV and $\sigma_{RV}$ of the MW bulge in the region $|l,b|\leq10^\circ$. 
These maps, expressed as functions of Galactic coordinates, were derived by interpolating data from 
30 observed fields across the entire bulge region, assuming two\---folds symmetry.
The GIBS fields, each containing between 100 and 400 red clump (RC) giants, extend as close as 
$\approx$2 degrees from the Galactic plane (see Fig.\ref{fig:old_sigma}).
The RV map revealed clear signature of cylindrical rotation \--- namely, stars at different 
latitudes exhibit similar RV values at a given longitude, similar results
also found by \citet{howard+2009, ness+2013kin_argosIV} among other authors.
In contrast, the $\sigma_{RV}$ map showed a prominent central peak, reaching values around 
140\,km/s and extending along 2 degrees from the center, which is equivalent to be
confined within a radius of $\approx 280$\, pc, considering each degree as $140$ pc 
assuming a distance to the center of 8 kpc.
Accurately characterizing this central peak, particularly in terms of its spatial extension, 
is crucial, as it may be linked either to a central stellar density enhancement 
\citep{valenti+2016} that could be related to an inner component like the nuclear star cluster 
\---which hosts a total stellar mass of $\sim3\times10^7M_{\odot}$ in a small radius of $\sim4$~pc, 
see \citet{neumayer+20} for further details\---, or to the presence of elongated 
orbits along the line of sight, often 
referred to as anisotropy of the orbits \citep{simion+21}.
However, due to the challenges posed by high extinction and stellar crowding, the innermost 
regions near the Galactic plane remain poorly constrained.
To address this limitation, \citet[][hereafter V18]{valenti+2018} supplemented the GIBS kinematic 
maps with four additional inner fields ($|b|\sim2^\circ$) observed with the integral field unit 
(IFU) spectrograph MUSE \citep{MUSE} at the ESO Very Large Telescope (VLT). 
Despite the low resolution of MUSE ($R\approx3000$), the RV was measurable by means of the prominent absorption lines of the Calcium II triplet 
(hereafter CaT; an example of these lines around wavelength $8600 \AA$ is shown in Fig.~\ref{fig:specs}).
These new observations, based on significantly larger stellar sample (for instance, $\approx 500-1200$
per field), confirmed both the vertical extent of the $\sigma_{RV}$ peak at positive
latitudes and its symmetry with respect to the Galactic plane. 

To further constrain the kinematic properties of the stellar population in the inner bulge, we
present an analysis of nine newly observed fields using the MUSE spectrograph, employing 
similar methodology as in V18. 
In addition, by integrating archival data from V18, GIBS and APOGEE, we update
the existing maps of mean RV and $\sigma_{RV}$ maps, deriving revised analytical expressions
as functions of Galactic coordinates. 
Finally, the new MUSE data were processed using the {\tt PHOTfun} code, which performs standard
Point Spread Function (PSF) photometry on the monochromatic images obtained from the MUSE datacubes. 
Its {\tt PHOTcube} extension enables the extraction of stellar spectra by concatenating the flux 
measurements (derived from the magnitude conversion) of each detected source across the datacube. 
While {\tt PHOTfun} was originally designed for PSF photometry on arbitrary image sets, the 
integration of {\tt PHOTcube} extends its functionality to IFU datacubes, facilitating 
simultaneous photometric and spectroscopic analysis.
The complete toolkit is released as unified software suite under the name {\tt PHOTfun}.

\section{The spectroscopic sample}  
\label{sec:obs}

The kinematic maps presented in this study were derived by combining several different 
spectroscopic datasets, some of them
analyzed here for the fist time, and some other already published. In total, the maps are based
upon $\sim$23000 stars,
spread across 57 bulge fields. We describe each datasets here below. 

\smallskip
\paragraph{MUSE-inner.} These are three fields located within $\sim$\,150\,pc from the Galactic center, observed within ESO proposal 101.B-0381 (P.I. Zoccali, M). The original purpose 
of these data was to characterize the $\sigma_{\rm RV}$ peak, in the inner 
bulge, presented for the first time by Z14. The reduction of these data are described 
here for the first time, in Sec.~\ref{sec:muse}. Table\,\ref{tab:MuseObs} lists 
the Galactic coordinates, the exposure time, the image quality and extinction 
\citep{surot+20redd} of all MUSE fields used in this work.

\smallskip
\paragraph{MUSE-outer.} Six additional fields were observed further away from the Galactic center,
within ESO proposal 101.D-0363 (P.I. Minniti, D). The fields were centered at the coordinates of candidate new
globular clusters, later discarded by \citet{gran+19}. Therefore, they sample bulge field (and
foreground disk) stars, and had much shorter exposure times. The reduction and analysis of 
these data are also described here for the first time (Sec.~\ref{sec:muse}). 

As these datasets (MUSE-inner and -outer) are being for the first time 
analyzed, we refer to them as the new MUSE data. Each field name follows the same codename convention as the GIBS data, based on the Galactic coordinates; for example, (l$^\circ$=0.2, b$^\circ$=1.3) is labeled as p0.2p1.3.

\smallskip
\paragraph{MUSE-V18.} These are the four fields in the inner bulge, acquired within ESO proposal 99.B-0311 and discussed in V18. We use here directly the RV measurements obtained in that paper. 

\smallskip
\paragraph{GIBS data.} The main goal of the present work is to update the kinematic maps presented 
by Z14. Therefore, we included here all the data from that paper. They consist in RV 
measurements for a sample of $\sim$5000 RC stars only, spread across 30 fields, mostly at
negative latitudes. The spectra were obtained with the FLAMES-GIRAFFE\footnote{Based on
observations taken with ESO telescopes at the La Silla
Paranal Observatory under ESO programme IDs 187.B-909 and 089.B-0830} spectrograph \citep{FLAMES}, at resolution
R$\sim$6500, and were centered on the infrared CaT region at $\sim$8500\AA. 

\begin{figure}

	\includegraphics[width=\hsize]{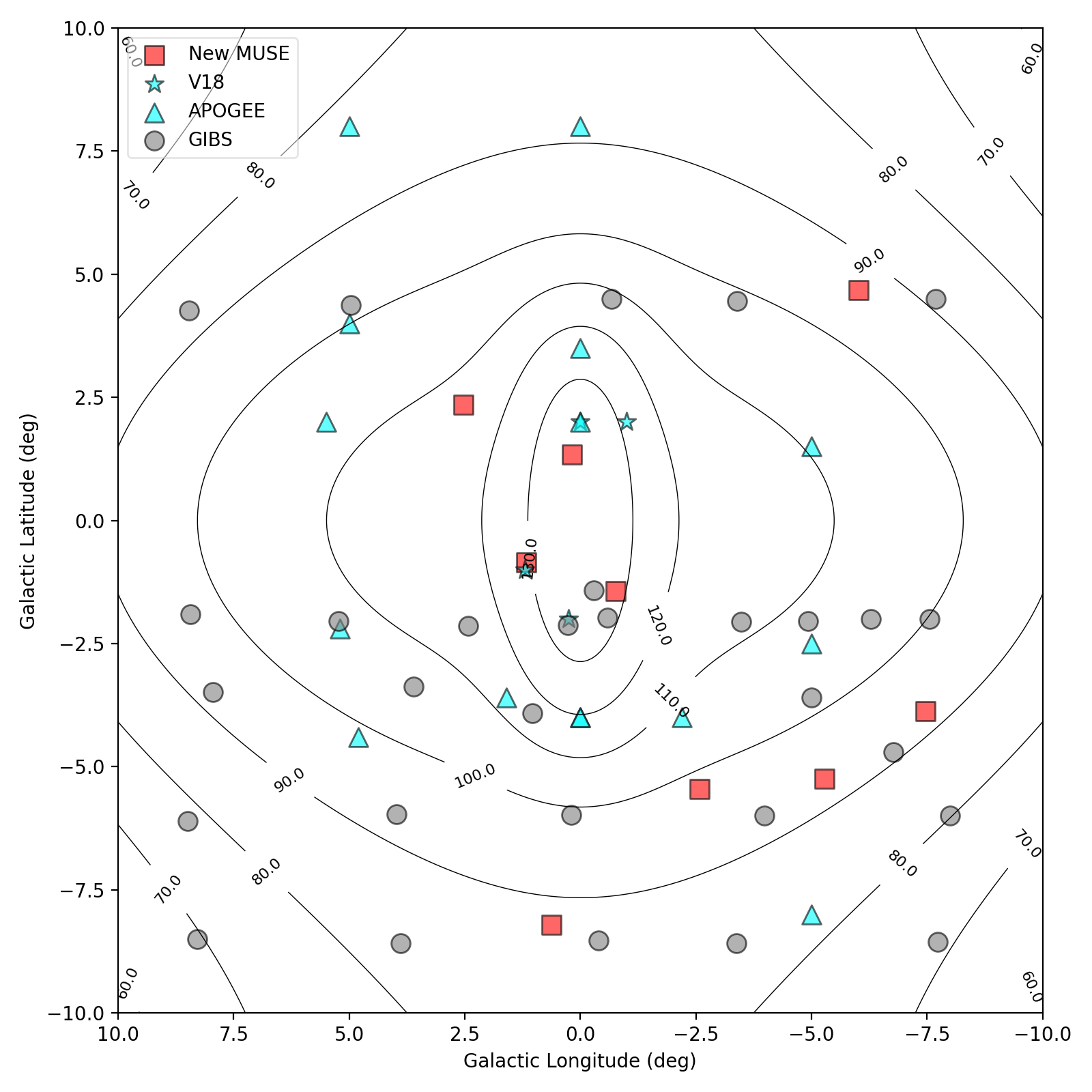}
 \caption{Spatial distribution in Galactic coordinates of the global spectroscopic sample considered in this study compared to the $\sigma_{RV}$ contours (solid lines) of the Z14 map. The GIBS fields used to derive the map of Z14 are marked as gray circles, while the cyan symbols refer to the additional fields used in this work to derive new map. Specifically, the new MUSE fields (red squares), the APOGEE fields (cyan triangles), and the V18 fields (cyan stars). }
    \label{fig:old_sigma}
\end{figure}

\smallskip
\paragraph{APOGEE data.} The APOGEE survey \citep{apogee} targeted red giant stars across 
the whole MW. For the present study, we selected bona\---fide bulge stars from the DR17 
catalog \citep{apogee-DR17} by following the prescriptions of \cite{rojas-arriagada+2020}.
Briefly, we retained all M giants in the region between $|l|\leq10^\circ$ and 
$|b|\leq10^\circ$, with spectro\---photometric distance within 3.5\,kpc from the Galactic 
center. Additionally, to account for metallicity bias due to the survey selection function 
and to the limited sampling of the models grid of the APOGEE pipeline, we further excluded 
all stars with log\,(g)$\ge$2.2 \citep[see Sec.~3.1 in][]{rojas-arriagada+2020}. 
The so\---derived APOGEE catalog has been used to further select 14 bulge regions with a
relatively high target density (for example, $\ge$100 stars per circular area of 1 degree of 
radius), and derived a RV and $\sigma_{RV}$ for each of them.
\\
\\
Finally, it is worth stressing that the selection of all spectroscopic fields, both new and
previously published, aims at enhancing the  spatial sampling in the region between 
$|l|\leq10^\circ$ and $|b|\leq10^\circ$, and hence enabling a more precise constraint on 
the existing map, especially in the innermost regions (see Fig.\,\ref{fig:old_sigma}).

\section{Reduction and analysis of the new MUSE data}
\label{sec:muse}

For the new datasets, MUSE-inner and MUSE-outer, we provide here a detailed description of the observations, data reduction, spectrum extraction and RV measurements.

\subsection{Observations and data reduction}
\label{subsec:muse_obs}
As part of an ongoing project aimed at probing the kinematics and chemistry of the stellar population of the inner Galactic bulge, we have collected deep MUSE observations of three 
fields (for instance, p0.2p1.3, m0.7m1.4, p1.2m0.9) within $\sim$150\,pc from the Galactic center.
We have carried out multiple MUSE adjacent pointing leading to a mapped area of about 3\,arcmin$^2$ 
per field, and performed multiple visits to enable the detection and RV measurement
of dwarfs $\sim$1 mag fainter than the bulge main\---sequence turnoff (for instance, I$\sim$19), 
with signal-to-noise (SNR)$\ge$12.
Six additional fields, consisting of just one pointing each, have been found in the ESO/MUSE 
archive as part of a program aimed at searching for obscure bulge clusters, and have been 
included in our sample to improve the spatial coverage across the bulge (see Table.\ref{tab:MuseObs}).

MUSE \citep{MUSE} consists of 24 identical IFU that \--- when used as in our case in Wide Field Mode (WFM) \--- together sample a nearly contiguous area of 1 arcmin$^2$, with a spatial resolution of 0.2"/px.
The observations of all fields have been combined with the Ground Layer Adaptive Optics mode (for instance, WFM-AO) of the VLT Adaptive Optics Facility \citep{VLT-AOF} through the GALACSI AO module \citep{GALACSI}, which acts as seeing enhancer.
By using the so-called Nominal setup (for instance, WFM-AO-N), the resulting spectra span mostly 
over the whole optical range \--- for instance, from 4800\,$\AA$ to 9300\,$\AA$, with a gap 
5820\---5970\,$\AA$ due to the Na Notch filter blocking the emission from the lasers 
\--- with a resolution of R$\sim$3100 at 8000\,$\AA$, and a sampling of 1.25\,$\AA$.
For all the fields, similar observing strategy but different total integration time 
(see Table\,\ref{tab:MuseObs}) has been used: a combination of on\---target sub\---exposures, 
taken with a small offsets pattern (for instance, $\sim$1.5") and 90$^\circ$ rotations to optimize 
the cosmic rays rejection and obtaining a uniform combined dataset in terms of noise properties.

Raw data processing has been carried out by using the MUSE pipeline \citep{MUSEpipeline}. As first step, the pipeline produces the master calibrations needed to account for the instrumental effects (for example, bias, flats, arc lamps, line spread function, illumination, geometrical distortion and response curve for flux calibration).
Subsequently, the correction by using the master calibrations is performed for each exposure of each
individual IFU, and a pixel table containing the processed data coming from all the 24 IFUs is created.
For each MUSE pointing, a final datacube is then obtained by combining together the pixel tables corresponding to all the available exposures.
Additionally, we have used the pipeline option of producing the so-called field\---of\---view (FoV) images by convolving the MUSE datacube with the transmission curve of various filters. 
For this work we produced FoV images in V\---Johnson, R\---Cousins, and I\---Cousins, as well as the white light FoV image covering the whole spectral range.

\begin{table}
\caption{Parameters of the new MUSE observations.}
\label{tab:MuseObs}      
\centering
\tiny
\begin{tabular}{l c  c c c c}
\hline\hline
Field  &  l  &  b  &   Exp. Time &  FWHM & E(J\---K)  \\   
& [deg] & [deg] &  [sec] & ["] & mag\\
\hline
&&&&&\\
&0.19 &1.30 &  6$\times$1100 & 0.97 &0.920\\
p0.2p1.3&0.18 &1.34 &  6$\times$1000 & 0.76 &0.885\\
        &0.19 &1.35 &  6$\times$1000 & 0.77 &0.998\\
&&&&&\\
 &-0.75 &-1.42 &  6$\times$1000 & 0.80 &0.991\\
m0.7m1.4&-0.77 &-1.43 &  6$\times$1100 & 0.92 &1.019\\
 &-0.79 &-1.44 &  6$\times$1020 & 1.00 &0.996\\
 &&&&&\\
 &1.17 & -0.87 & 6$\times$1000 & 0.92 &0.851\\
p1.2m0.9  &1.16 & -0.85 &6$\times$1000 & 0.85 &0.878  \\
 & 1.17 & -0.84 & 1$\times$1000 & 1.02 &0.744\\
&&&&&\\
m5m5 &-5.28 &-5.25 & 3$\times$1035 & 0.81 &0.217\\
p1m8 &0.62 &-8.22 & 3$\times$700 & 0.65 & 0.102\\
p3p2 &2.53 &2.34 & 3$\times$1035 & 0.61 &0.623\\
m7m4 & -7.46&-3.87 & 3$\times$1035 & 0.79 &0.472 \\
m3m5 & -2.58 &-5.46 & 3$\times$1035 & 0.59 & 0.255\\
m6p5 & -6.02&4.68 & 3$\times$1030 & 0.72 &0.341\\
\hline  
\end{tabular}
\tablefoot{Columns list names, galactic coordinates, exposure
times, image quality as measured on the FoV images, reddening from \citet{surot+20redd}.
}
\end{table}

\subsection{Software release: PHOTfun and PHOTcube}
\label{sec:PHOTsuite}

{\tt PHOTfun} is a Python package developed in the context of this work to facilitate PSF 
photometry tasks based on the well established software packages DAOPHOT-II and ALLSTAR 
\citep{daophot}. It also includes a dedicated extension, {\tt PHOTcube}, specifically designed 
for spectral extraction from IFU datacubes. In crowded stellar fields, spectral extraction 
requires PSF-based photometry to accurately deblend sources. While previous tools such 
as PampelMUSE \citep{PampelMuse} implement similar functionality, they present key limitations: (i) they 
require a predefined list of targets for spectral extraction, (ii) they do not allow 
interactive selection of the stars used for PSF modeling, and (iii) they are less effective 
at subtracting sky artifacts compared to DAOPHOT-II–based methods.
The {\tt PHOTfun} package consists of two main components: a graphical user interface (GUI) 
that streamlines the use of DAOPHOT-II for photometry, and the {\tt PHOTcube} extension, 
which enables efficient and scalable spectral extraction from IFU datacubes.

{\tt PHOTfun} in particular is a GUI designed to simplify the use of DAOPHOT-II, an example of the main menu is shown in the appendix (see Fig.~\ref{fig:photfun_main}). The GUI 
is based on the python package for web apps Shiny \citep{shiny}. It works on every set of 
images and includes the DAOPHOT-II subroutines FIND, PICK, PHOT, PSF, SUBTRACT and 
DAOMATCH that are necessary for detecting the source, finding their coordinates on every image in the given set, modeling the PSF and finally performing the photometry. This last step is done using the ALLSTAR standalone software included in DAOPHOT-II. 
While {\tt PhotFun} itself does not add new capabilities to DAOPHOT-II, 
it is a user-friendly interface that can be used to perform point-source detection 
and PSF photometry on any sets of astronomical images, visualizing the most important 
intermediate results and interactively rejecting model PSF stars as it is shown in
Fig.~\ref{fig:photfun_selection}. 
Interoperability via Simple Application
Messaging Protocol (SAMP) to work with software like TOPCAT \citep{topcat} 
and DS9 \citep{ds9} is also enabled.

{\tt PHOTcube} is an extension integrated into the {\tt PHOTfun} GUI, allowing users to load 
the datacube and slice it along the wavelength direction in a set of sequential monochromatic images that can be processed 
through the {\tt PHOTfun} GUI by means of standard PSF photometry using DAOPHOT-II. 
Finally, for each detected source, the {\tt PHOTcube} extension compiles the monochromatic magnitudes, converts them into fluxes and generates the corresponding spectra.

The full software and source code is available at
Github\footnote{https://github.com/ciquezada/photfun} and 
PyPi\footnote{https://pypi.org/project/photfun/}. To enable the use of 
DAOPHOT, {\tt PHOTfun} leverages a Docker\footnote{https://www.docker.com/products/docker-desktop/}
container, which must be installed to ensure compatibility 
and ease of use across different systems.

\subsection{Spectrum extraction: Step by step}
\label{sec:extraction}

The extraction of stellar spectra from the MUSE datacubes using {\tt PHOTfun} followed 
these steps: a master target list for each field was generated using the FIND routine. The FoV image corresponding to the white light, which integrates the 
total flux across all wavelengths, was used for optimal source detection. For PSF 
modeling, the PICK routine performs an automatic pre-selection of suitable stars; 
however, it is recommended to manually inspect and refine this selection using 
the GUI tool, which allows a visual inspection of the light profile of each star 
(see Fig.~\ref{fig:photfun_selection}). 
At this stage the master and PSF target lists were defined for each datacube.

The {\tt PHOTcube} extension slices the datacube and loads it as a sequence of monochromatic 
images into {\tt PHOTfun}. Using the defined master and PSF target lists, PSF photometry 
was performed on each slice with the ALLSTAR routine, measuring the monochromatic 
fluxes for each target. The individual spectra were then reconstructed by concatenating 
the measured fluxes across the full wavelength range. This method had been used before by,
for example, V18, \citet{olivares22}. The signal\---to\---noise ratio (SNR) 
of each spectrum were
computed as the mean SNR per wavelength pixel, where the SNR of each pixel were defined 
as the ratio of flux to its associated uncertainty.

\begin{figure}

	\includegraphics[width=\hsize]{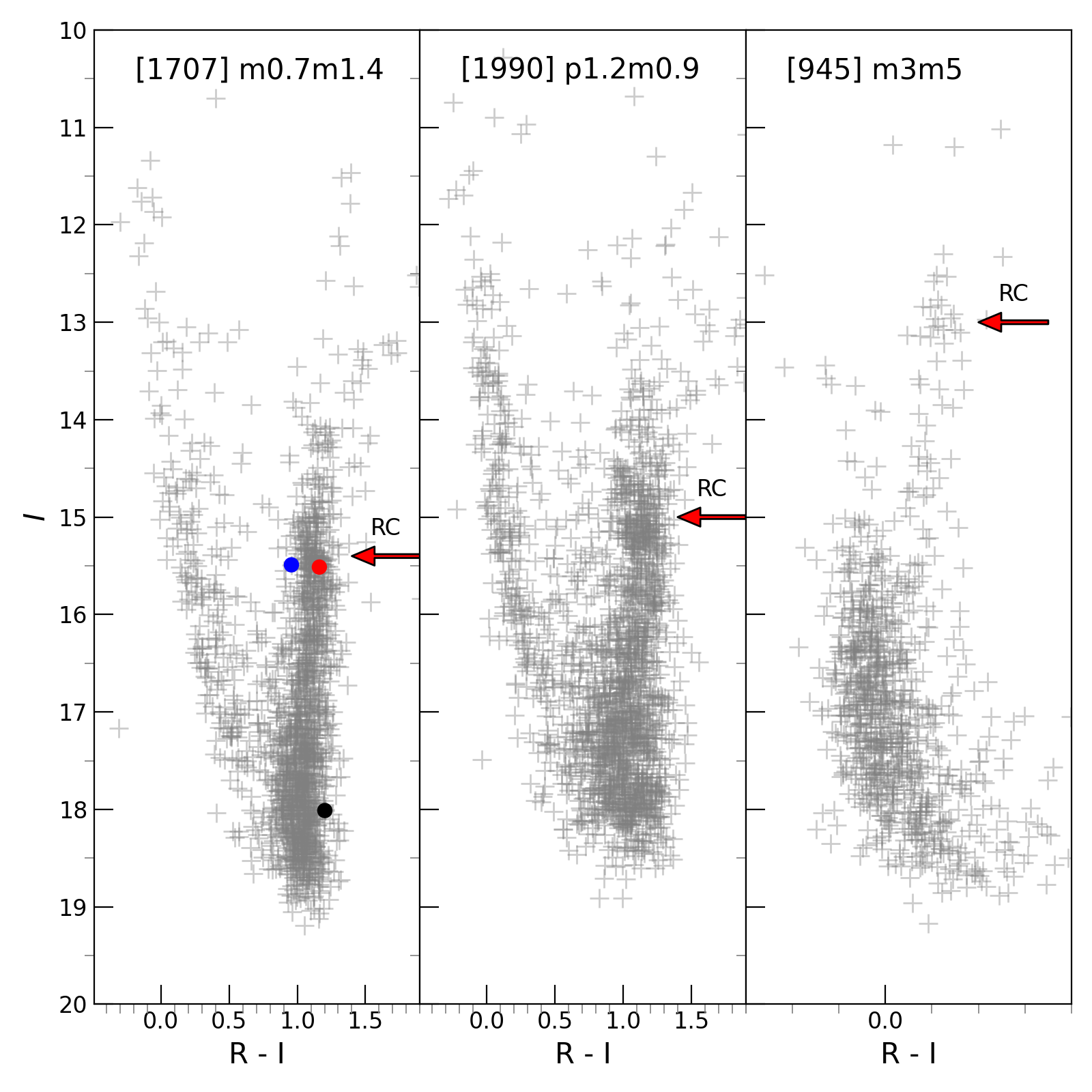}
 \caption{CMD of three MUSE fields obtained by performing PSF\---fitting photometry on the FoV images. The names of the fields and number of detected stars are given inside at the top of each panel. The red arrow marks the position of the RC in each CMD. Also shown in the left panel shows the position of three stars for which we show the extracted spectra in Fig.~\ref{fig:specs}.}
    \label{fig:cmds}
\end{figure}

Additionally, using {\tt PHOTfun}, a color–magnitude diagram (CMD) was 
produced for each field. 
This was done by performing PSF photometry on the MUSE pipeline FoV 
images using the V\---Johnson, 
R\---Cousins, and I\---Cousins filters, and the same master and PSF target lists. 
Fig.~\ref{fig:cmds} shows an example of the derived CMD of three fields using the measured
instrumental magnitude, two from the 
MUSE-inner sample and one from the MUSE-outer sample (see section~\ref{subsec:muse_obs}). 
Notable differences in the CMD statistics and photometric depth arise from variations 
in the field extinction, coverage and integration times across the sample (for instance, Table.\,\ref{tab:MuseObs}).  

The CMDs of the observed fields allowed for partial separation of the bulge population from 
the foreground disk stars along the line of sight. Specifically, the disk main sequence is 
clearly visible as a narrow, vertical blue sequence that merges with the bulge population 
near the main sequence turnoff (MS-TO). For an old stellar population, the MS-TO is 
typically located approximately 3.2 magnitudes fainter than the red clump (RC) in the 
I band \citep{BASTIiso}.

\begin{figure}

	\includegraphics[width=\hsize]{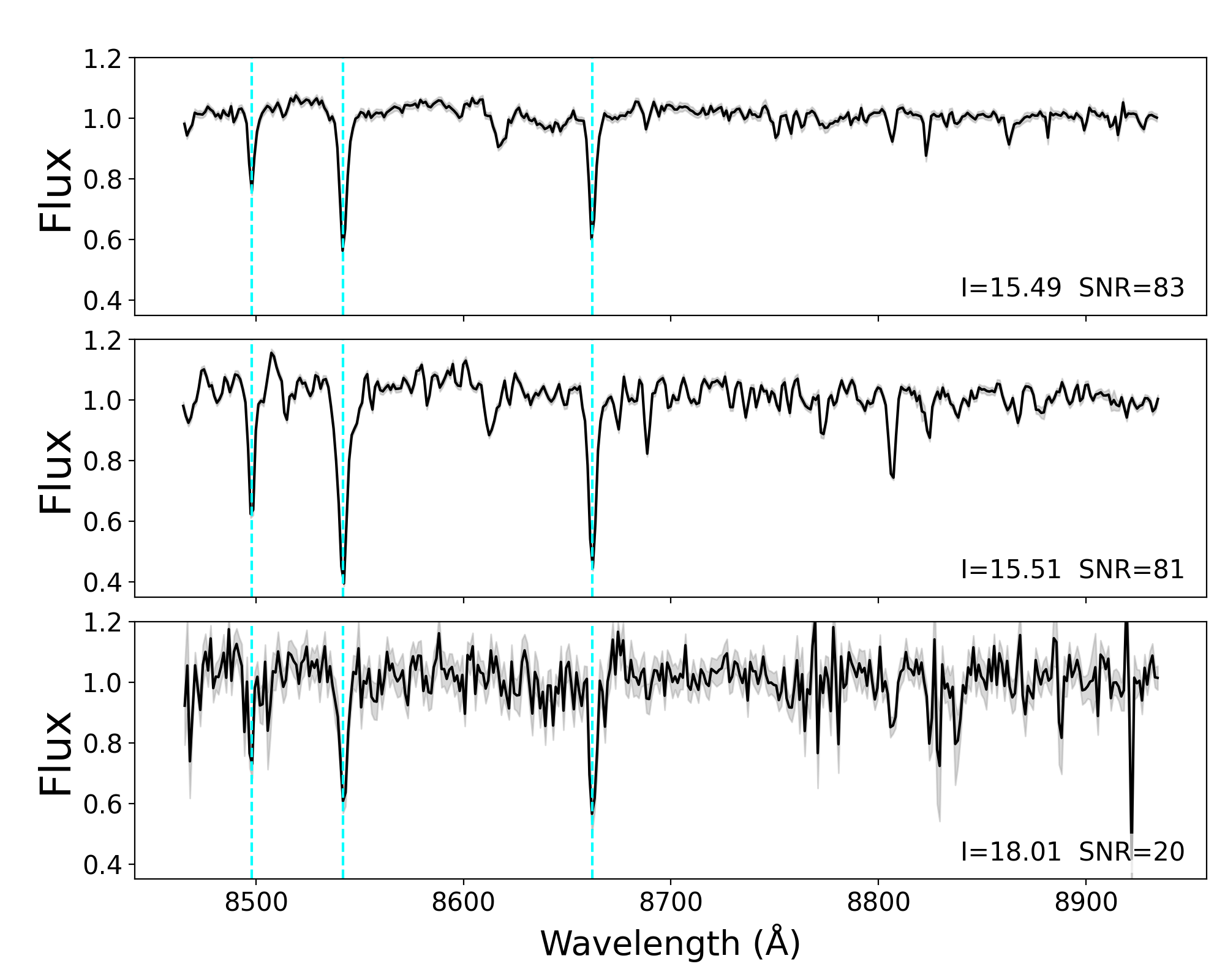}
 \caption{CaT region of the normalized spectra (Sec~\ref{sec:extraction}) for three stars within the field m0.7m1.4. Vertical cyan lines mark the position of the three CaT lines. The top and middle panels show the two stars marked as blue (top) and red (bottom) dots in the CMD shown in the left panel of Fig.~\ref{fig:cmds}. Their position in the CMD and the strenght of their CaT lines suggest that they are metal-poor and a metal-rich star, respectively. 
 The bottom panel shows the faint star marked as a black dot in the left panel of Fig.~\ref{fig:cmds}. A gray shaded area shows the flux error in the three panels, as estimated from the magnitude errors provided by DAOPHOT. Due to the different SNR of the three spectra (shown in the figure labels), the gray shaded area is only visible in the bottom panel.
}
    \label{fig:specs}
\end{figure}

Table~\ref{tab:RVs} summarizes the number of stellar spectra extracted for each field,
while Fig.~\ref{fig:specs} presents representative examples from field m0.7m1.4, highlighting the spectral region around the CaT. 
The selected spectra correspond to stars of varying metallicity, identified based on their position along the red giant branch in the CMD, where redder stars generally indicate higher metallicity. This trend is evident in the middle panel, where the redder star exhibits noticeably stronger CaT absorption features. 
For comparison, a low SNR spectrum is also shown, illustrating the increased uncertainty in such cases. 
Nevertheless, all spectra clearly display the CaT absorption lines, which are critical for precise radial velocity and metallicity determinations.

\begin{table}[!]
\caption{Fields characteristics}.
\label{tab:RVs}      
\centering
\tiny
\begin{tabular}{l c c c c }
\hline\hline
Field     &  N  &  $\mu RV_{hc}$  &  $\sigma RV$  &  RMSE \\   
& [stars] & [km/s] & [km/s] & [stars] \\
\hline
&&&&\\ 
\textbf{MUSE-inner} &&&\\
p0.2p1.3   &   1873  &    -1 $\pm$ 6  &   134 $\pm$ 5  &    11   \\
m0.7m1.4   &   1707  &   -21 $\pm$ 7  &   137 $\pm$ 7  &    13   \\
p1.2m0.9   &   1990  &     2 $\pm$ 6  &   129 $\pm$ 5  &    13   \\
&&&\\ 
\textbf{MUSE-outer} &&\\
m5m5       &   542  &   -68 $\pm$ 8  &    94 $\pm$ 6   &   6   \\
p1m8       &   386  &    -3 $\pm$ 9  &    90 $\pm$ 6   &   4   \\
p3p2       &    735  &    30 $\pm$ 9  &   126 $\pm$ 7  &   17   \\
m7m4       &   577  &   -78 $\pm$ 8  &    91 $\pm$ 6   &   5   \\
m3m5       &   945  &   -43 $\pm$ 7  &   100 $\pm$ 5   &   5   \\
m6p5       &    605  &   -78 $\pm$ 8  &    94 $\pm$ 6  &   6   \\
&&&\\ 
\textbf{MUSE-V18} &&&\\
p0m2         &     1203  &     9  $\pm$ 4 &    135 $\pm$  3 &  -  \\ 
m1p2         &     861  &    -8  $\pm$ 5 &    125 $\pm$  3  &  -  \\ 
p0p2         &     496  &    19  $\pm$ 8 &    137 $\pm$  5  &  -  \\ 
p1m1         &     502  &    14  $\pm$ 7 &    119 $\pm$  5  &  -  \\ 
&&&\\ 
\textbf{APOGEE} &&&\\
m5m2.5     &     197   &   -49 $\pm$ 14 &   105 $\pm$ 10 &  8 \\
m5p1.5     &     132   &   -66 $\pm$ 17 &   103 $\pm$ 12 &  12 \\
p0m4       &     348   &    -0 $\pm$ 12 &   122 $\pm$  9 &  7 \\
p0p8       &     304   &    -8 $\pm$ 10 &    90 $\pm$  7 &  9 \\
p0p3.5     &     439   &    -8 $\pm$ 11 &   120 $\pm$  7 &  13 \\
p0p2       &     416   &     3 $\pm$ 13 &   135 $\pm$  9 &  10 \\
p5.5p2     &     490   &    58 $\pm$  8 &    99 $\pm$  6 &  25 \\
p5p4       &     391   &    52 $\pm$ 10 &   100 $\pm$  6 &  16 \\
p5.2m2.2   &     563   &    57 $\pm$  8 &   105 $\pm$  6 &  30 \\
p4.8m4.4   &     291   &    50 $\pm$ 11 &    99 $\pm$  8 &  7 \\
p1.6m3.6   &     429   &    22 $\pm$ 10 &   114 $\pm$  7 &  14 \\
m2.2m4     &     330   &   -36 $\pm$ 12 &   118 $\pm$  9 &  5 \\
m5m8       &     200   &   -53 $\pm$ 11 &    84 $\pm$  8 &  4 \\
p5p8       &     189   &    28 $\pm$ 10 &    73 $\pm$  7 &  3 \\
&&&\\ 
\textbf{GIBS} &&&\\
p8.4p4.3         &     209  &     69 $\pm$ 11 &    86 $\pm$  8 &   5  \\ 
p4.3p4.4         &     208  &     50 $\pm$ 13 &    98 $\pm$  9 &   5  \\ 
m0.7p4.5         &     210  &      1 $\pm$ 15 &   112 $\pm$ 10 &   8 \\ 
m7.7p4.5         &     210  &    -70 $\pm$ 12 &    93 $\pm$  9 &   6  \\ 
m3.4p4.5         &     208  &    -64 $\pm$ 13 &   100 $\pm$  9 &   7  \\ 
m0.3m1.4         &     441  &     -5 $\pm$ 12 &   135 $\pm$  8 &   10  \\ 
m7.5m2.0         &     210  &    -86 $\pm$ 13 &    96 $\pm$  9 &   5  \\ 
m6.3m2.0         &     113  &    -54 $\pm$ 19 &   102 $\pm$ 13 &   14 \\ 
m4.9m2.0         &     209  &    -43 $\pm$ 14 &   108 $\pm$ 10 &   7 \\ 
m3.5m2.1         &     111  &    -58 $\pm$ 21 &   116 $\pm$ 15 &   16 \\ 
m0.6m2           &     111  &    -16 $\pm$ 25 &   137 $\pm$ 18 &   21 \\ 
p8.4m1.9         &     209  &     70 $\pm$ 13 &    95 $\pm$  9 &   5  \\ 
p5.2m2.1         &     112  &     80 $\pm$ 20 &   111 $\pm$ 14 &   15 \\ 
p2.4m2.1         &     318  &     57 $\pm$ 12 &   115 $\pm$  9 &   4  \\ 
p0.3m2.1         &     435  &     12 $\pm$ 12 &   135 $\pm$  9 &   9  \\ 
p8m3.5           &     105  &     76 $\pm$ 17 &   89. $\pm$ 12 &   11 \\ 
p3.6m3.4         &      91  &     52 $\pm$ 23 &   112 $\pm$ 16 &   17 \\ 
m5.0m3.6         &     108  &    -61 $\pm$ 19 &   102 $\pm$ 13 &   14 \\ 
p1.0m3.9         &     102  &     22 $\pm$ 22 &   115 $\pm$ 16 &   16 \\ 
m6.8m4.7         &     107  &    -57 $\pm$ 16 &    84 $\pm$ 11 &   10 \\ 
p8.5m6.1         &     209  &     66 $\pm$ 10 &    77 $\pm$  7 &   5  \\ 
p4m6             &     213  &     43 $\pm$ 12 &    92 $\pm$  8 &   4  \\ 
p0.2m6           &     454  &      0 $\pm$  9 &    98 $\pm$  6 &   33  \\ 
m8m6             &     215  &    -71 $\pm$ 10 &    80 $\pm$  7 &   3  \\ 
m4m6             &     224  &    -55 $\pm$ 12 &    93 $\pm$  8 &   5  \\ 
p8.3m8.5         &     193  &     50 $\pm$ 11 &    79 $\pm$  7 &   3  \\ 
p3.9m8.6         &     207  &     21 $\pm$  9 &    68 $\pm$  6 &   6  \\ 
m0.4m8.5         &     415  &     -8 $\pm$  8 &    89 $\pm$  6 &   23  \\ 
m7.7m8.5         &     206  &    -71 $\pm$ 11 &    81 $\pm$  8 &   4  \\ 
m3.4m8.6         &     208  &    -32 $\pm$ 10 &    79 $\pm$  7 &   6  \\

\hline  
\end{tabular}
\tablefoot{Number of stars adopted in each bulge field, mean RV, $\sigma_{RV}$ and RMSE. RMSE for New MUSE fields correspond to the double Gaussian fitting. V18 RMSE are missing because they are not reported in the original paper}
\end{table}

As previously mentioned, another widely used tool for spectral extraction from IFU data is PampelMuse \citep{PampelMuse}, which, in contrast to {\tt PHOTfun}, requires an external input catalog of sources. 
Another key distinction between the two methods lies in the approach to sky subtraction. 
While PampelMuse relies on the global sky subtraction provided by the MUSE pipeline, {\tt PHOTfun} \--- via DAOPHOT-II \--- computes a local sky estimate within an annulus around each detected source. 
This localized sky subtraction approach allowed our method to effectively mitigate residual sky features that were not fully corrected by the standard pipeline.
Furthermore, science fields exhibiting significant sky variability \---- such as star-forming regions \---- greatly benefit from the ability to measure and subtract the local sky background.
Finally, DAOPHOT-II is optimized to perform PSF photometry in crowded fields, thus mitigating the problem of blending. Visual inspection also helped
discarding blends, as two stars with different RVs show broader or double CaT lines in their spectra.

\begin{figure}

	\includegraphics[width=\hsize]{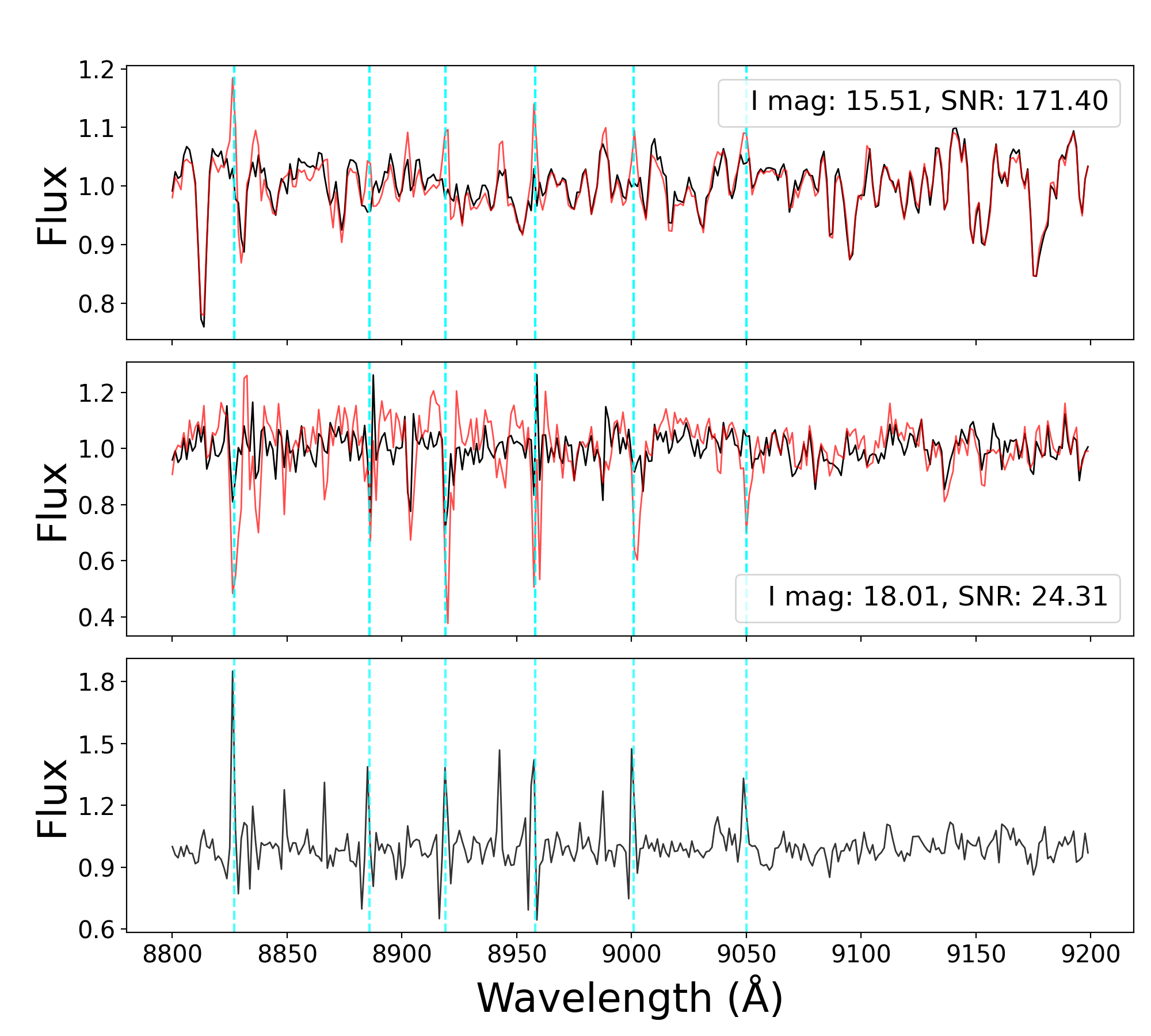}
 \caption{Comparison with PampelMuse. The top and middle panels show the normalized spectra (Sec~\ref{sec:extraction}) of a bright and a faint star respectively, as extracted by {\tt PHOTfun} in black, and red by PampelMuse. Both stars belong to the field m0.7m1.4. Vertical dashed cyan lines show the position of strong OH sky lines. The bottom panel shows the spectrum of a region of the image devoid of stars. Because the MUSE pipeline performs the subtraction of a single master sky spectrum, what we are seeing in the bottom panel shows the residual of this subtraction. These residuals are larger around the strong OH sky lines, consequently also the PampelMuse-extracted spectra show non-negligible residuals.}
    \label{fig:pampelmuse}
\end{figure}

Figure~\ref{fig:pampelmuse} illustrates a comparison between spectra extracted using {\tt PHOTfun-PHOTcube} (black line) and PampelMuse  (red line) for a bright star (top panel), a faint star (middle panel), and an empty sky region (bottom panel). The bottom panel reveals residuals from imperfect sky subtraction around strong OH emission lines, highlighted by vertical cyan lines. 
These residuals were also present in the PampelMuse spectra, particularly around the OH features, whereas the spectra extracted with {\tt PHOTfun} are notably smoother in these regions, demonstrating the improved sky handling of our method.

\subsection{Radial velocities}

Individual star RVs along the line of sight were determined by measuring the Doppler shift of the spectra in 
the CaT wavelength range, from $8465\AA$ to $8935\AA$. Each spectrum was cropped to this 
region and the continuum were normalized by fitting a fourth-degree polynomial. The 
observed spectrum was then cross-correlated with a small grid 
corresponding to a subset of synthetic templates generated 
using TurboSpectrum \citep{alvarez+98}, together with MARCS model atmospheres \citep{marcs} 
and the Gaia-ESO Survey line list published in \citet{heiter+21}, this subset is not complete -it corresponds to a sparse grid- and it was just intended to obtain RV, 
nevertheless, covering the surface parameter 
range T$_{\rm eff}$=4000-6000 K, $\log$\,g=1.5-4.5 dex and 
[M/H]=$-$1-+1 dex. First, a cross-correlation was performed 
to obtain an initial guess of the RV using a synthetic spectrum randomly 
selected from the grid. The spectrum was then rest-frame corrected for the resulting 
RV guess, and a $\chi$$^2$ test was then done to find the best 
fitting synthetic spectrum for that particular star over the entire subset of synthetic spectra. 
A second iteration to find the final RV was then performed 
cross-correlating the observed spectrum with the 
synthetic one. No interpolation was done, as for RV measurements a perfect match 
between the observed and template spectra was not necessary. This step yielded the 
final value for the RV, together with the rest-frame corrected spectrum for each star.

We visually inspected the selected template for each star and the corresponding 
cross-correlation function (CCF) to discard spurious results, typically excluding 
10–30\% of the faintest stars from each field. The uncertainty in the RV was calculated using 
one of the two formulas presented in \citet{valenti+2018} based on simulated MUSE spectra 
and calibrated for their observed data for different SNR and metallicity. 
For this work, given the metallicity distribution of the bulge, we used the RV uncertainty of a solar-metallicity dwarf star ($0.0$ dex) which was the most representative case. This was given by

\begin{equation}
\label{eq:err_rv}
\ln(\epsilon_{RV}) = 4.624 - 1.023 \ln(S/N) - 0.159 [Fe/H] + 0.120 [Fe/H]^2.
\end{equation}

\begin{figure}

	\includegraphics[width=\hsize]{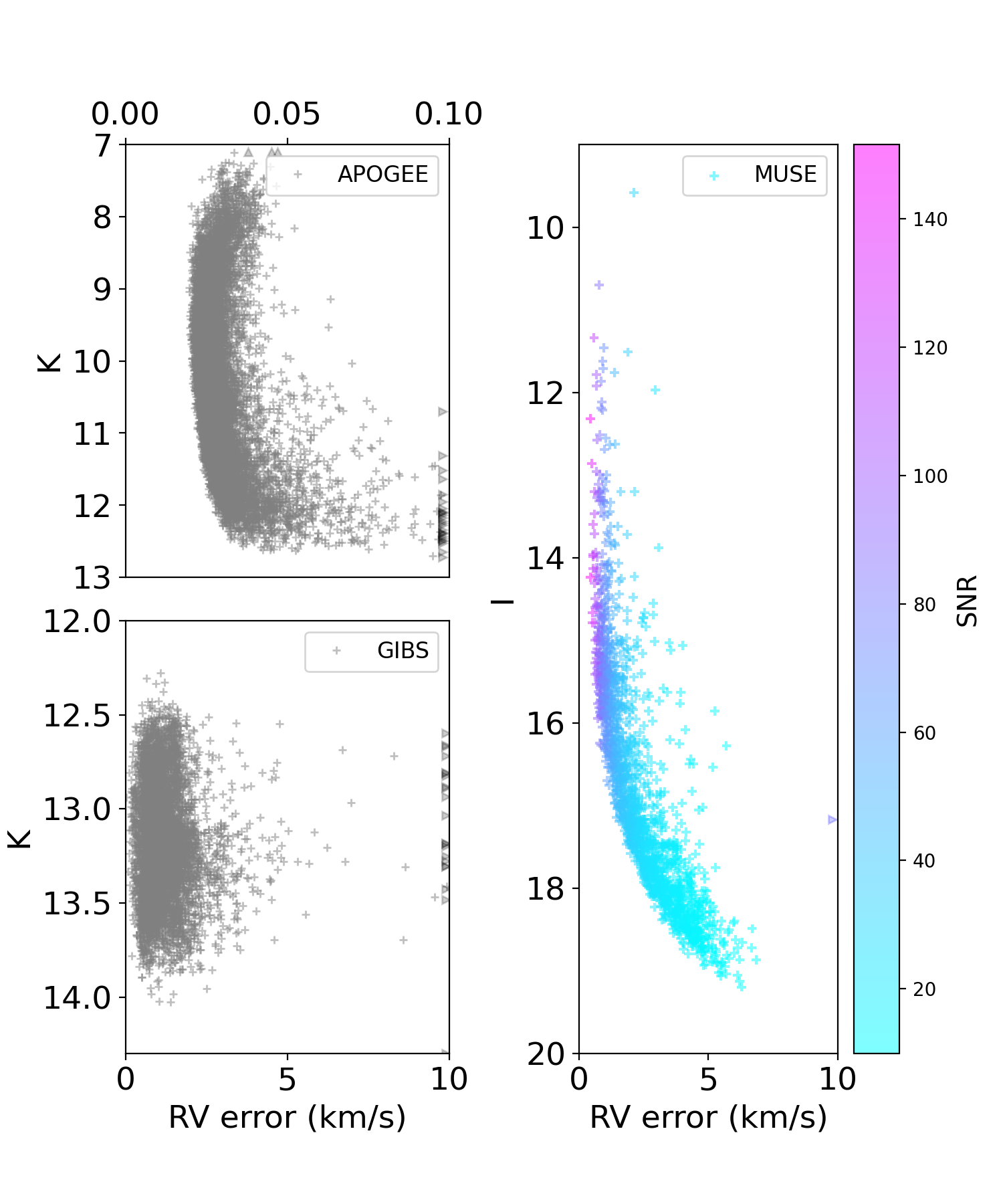}
 \caption{RV error versus magnitude for the APOGEE, GIBS, and MUSE samples. The MUSE example shown in the right-hand plot corresponds to the field m0.7m1.4 (see Fig~\ref{fig:cmds}). MUSE errors are estimated from the equation \ref{eq:err_rv}. Symbols show individual stars; arrows indicate values beyond axis limits. Note the differing axis scales: APOGEE RV errors are significantly smaller.}
    \label{fig:rv_err}
\end{figure}

An example of the derived uncertainty in RV is shown in Fig.~\ref{fig:rv_err}.
The derived RVs are already in the heliocentric reference frame, as this correction is performed by the MUSE pipeline. In order to transform them to a galactocentric reference frame, we use the formula
\citet[for example,][]{zoccali+14},

\begin{equation}
\label{eq:vcorr}
\begin{aligned}
V_{GC} &= V_{HC} + 220 \sin(l) \cos(b) \\
&\quad + 16.5 \left[ \sin(b) \sin(25) + \cos(b) \cos(25) \cos(l - 53) \right]
\end{aligned}
\end{equation}

\begin{figure}

	\includegraphics[width=\hsize]{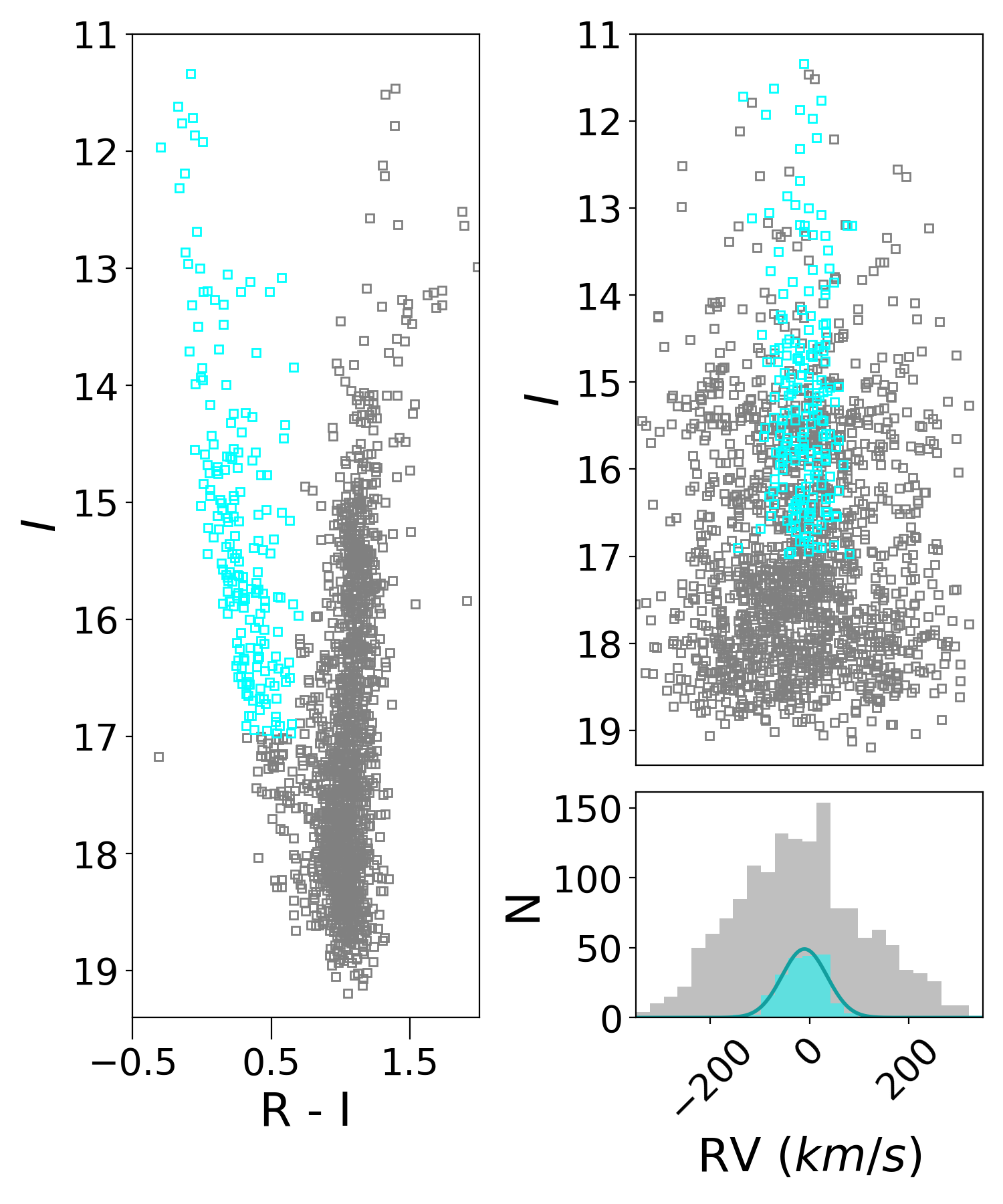}
 \caption{The left panel shows the CMD of the field m0.7m1.4. In cyan we have highlighted the disk subsample chosen to be a representation of the whole disk sample (in this field we chose to cut a magnitude $I<17$ 
and a color $R-I<0.8$). Top right panel shows the I magnitude vs the RV also with the whole disk sample highlighted. The bottom right panel shows the histogram with the highlighted sample of the disk}
    \label{fig:plot_cmd_disk}
\end{figure}

\begin{figure}

	\includegraphics[width=\hsize]{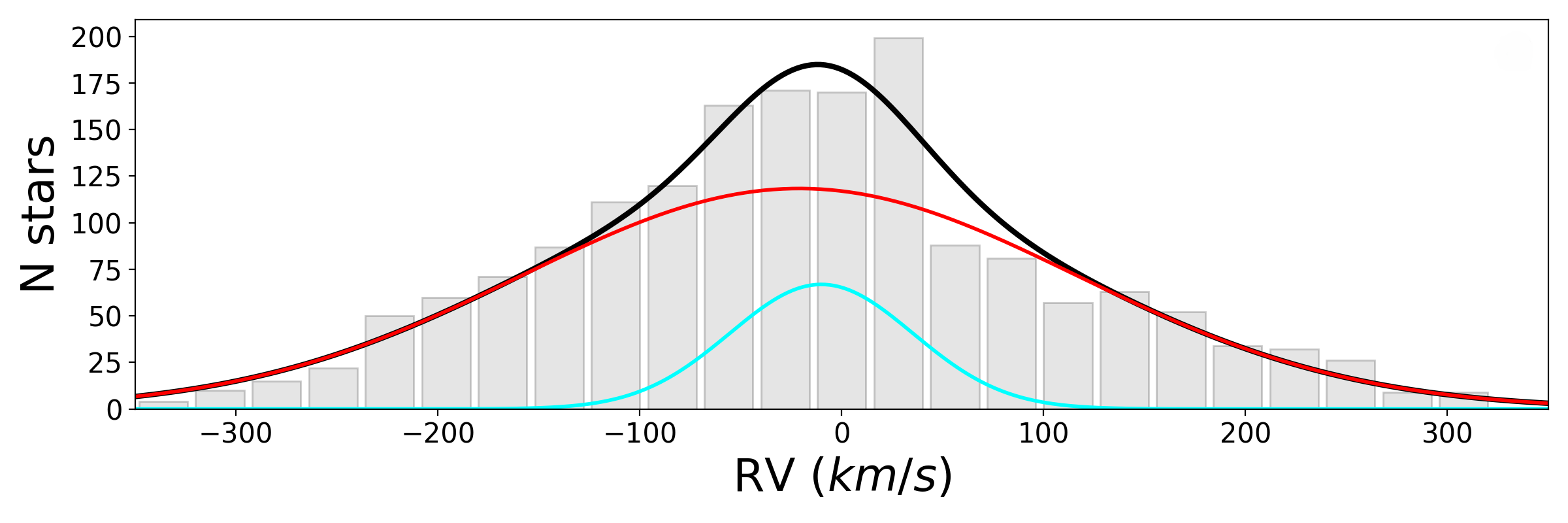}
 \caption{Fit of the two component gaussian profiles in the m0.7m1.4 field. The black line show the best fit that is a sum of the cyan profile that correspond to the disk distribution and the red profile that correspond to the bulge.}
    \label{fig:gaussian_fit}
\end{figure}

In order to derive the RV distribution of bonafide bulge stars in each field, we used the 
approach illustrated in Fig.~\ref{fig:plot_cmd_disk}. Each MUSE field samples a combination 
of bulge plus disk foreground stars. Based on the CMD,
we isolated bonafide disk stars in the bright blue sequence highlighted in cyan in 
Fig.~\ref{fig:plot_cmd_disk} (left). The top right panel of the same figure confirms that 
indeed, these stars were characterized by narrower RV distribution, compared to the total 
sample. The mean and the dispersion of this distribution were derived by a Gaussian fitting 
(bottom-right), and it was assumed that they represent the intrinsic RV and $\sigma_{RV}$ 
of the disk in this field. The RV and $\sigma_{RV}$ of the bulge was then 
derived by fitting the total observed distribution as the sum of two Gaussian profiles; 
the disk with the fixed parameters above, and the bulge with free parameters to be constrained. 
The latter fit was performed by means of the PyMC package \citep{pymc} which runs a Markov 
Chain Monte Carlo (MCMC) search for the best-fitting parameters. The best-fitting Gaussian 
profiles for the field m0.7m1.4 are shown in Fig.~\ref{fig:gaussian_fit}, where the 
red profile corresponds to the bulge which has mean RV$=-21$km/s and $\sigma_{RV}=137$km/s, and
the cyan one to the disk which has a mean RV$=-10$km/s and $\sigma_{RV}=45$km/s. The disk RV values
are comparable to the kinematic maps in \citet{ness+16kinAPOGEE}, Fig.7.

Concerning the literature data, we used the V18 RV and $\sigma_{RV}$ as they are provided 
in the paper. In contrast, for the GIBS and APOGEE data, where in each field 
we fitted a single Gaussian profile to the whole distribution of individual stars
galactocentric RV measurements provided in the catalogs, as they were already cleaned out of
foreground contamination. Indeed, GIBS selected targets were in a narrow magnitude range 
centered in the bulge RC, where disk contamination is negligible. For APOGEE data, the selection
of stars within 3.5 kpc from the galactic center already minimized foreground contamination.  
The RVs and $\sigma_{RV}$ for all the fields; the new MUSE data and the literature ones,
are listed in Table~\ref{tab:RVs}. We also report the root mean squared error (RMSE) of 
each individual and double Gaussian fit in Table~\ref{tab:RVs}. The RMSE values for the 
V18 fields are omitted since they were not reported in the original article. To avoid the low statistic in the fitting,
for fields with fewer than 150 stars we calculated the arithmetic mean RV and the associated
$\sigma_{RV}$ with the traditional formula. We obtained the same result as the Gaussian 
fitting with less than 1\% of difference; hence, we kept the values of the fitting 
for consistency.

\section{Kinematic maps}

\begin{table}
\caption{Coefficients of the kinematic maps for the RV and $\sigma_{RV}$.}
\label{tab:Coefs}      
\centering
\tiny
\begin{tabular}{l c c c c c c c c c c c c}
\hline\hline
&&&&&&&\\
\textbf{RV} &&&&&&&\\
\hline
        & A & B & C & D & E & F &  \\
Value   & -1.96  & -0.10  &  0.08  & 81.05  & -0.95  & 0.19   &    \\
Error   &  3.99  &  0.10  &  0.10  & 18.81  &  0.80  & 0.08   &    \\
&&&&&&&\\

&&&&&&&\\
\textbf{$\sigma$ RV} &&&&&&&\\
\hline
         & A & B & C & D & E & r & s \\
Value    & 45.54  & 53.36  & 281.61 & -0.12  & 36.30  &  693.96 & 21.60  \\
Error    & 11.96  & 13.68  &  38.47 &  0.06  &  4.87  &  169.46 &  6.41  \\
\hline 
\end{tabular}
\tablefoot{Derived from analytical equations described in Z14. The resulting best-fitting coefficients are listed along with their uncertainties}
\end{table}

The kinematic maps are updated by fitting phenomenological analytical equations previously used in Z14 to model the bulge RV and $\sigma_{RV}$, defined as follows:
\begin{align}
\mathrm{RV}_{GC} &= \left(A + B b^2\right) + \left(C b^2\right) l + \left(D + E b^2\right) \tanh(F l), \\
\sigma \mathrm{RV}_{GC} &= \left(A + B e^{-b^2/C}\right) + D l^2 + \left(E e^{-b^4/r}\right) e^{-l^2/s}.
\end{align}
The fitting process is performed using MCMC methods, following the same procedure as in the 
previous step, to derive the best-fitting coefficients. The resulting coefficients are 
presented in Table~\ref{tab:Coefs}, and the associated corner plots are shown in 
Fig.~\ref{fig:corner_rvgc} and Fig.\ref{fig:corner_sigmarvgc} in the appendix.

\begin{figure*}

	\includegraphics[width=\hsize]{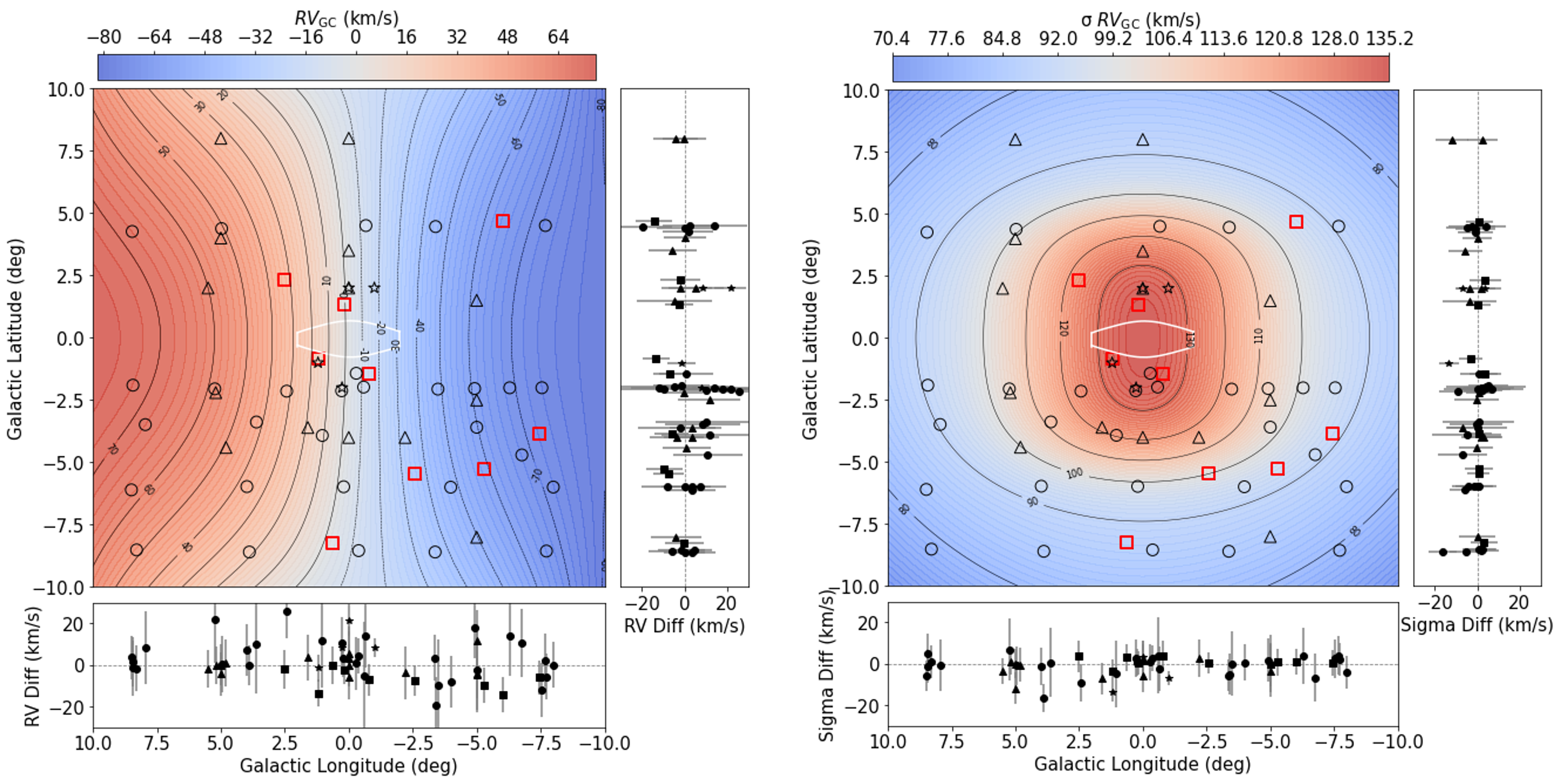}
 \caption{Updated maps of the bulge RV; colors and black contours show the RV profile fitted by the MCMC in steps of $10 \mathrm{km}/\mathrm{s}$. The different marks are showing the different source of the measurement following the same description as in Fig.~\ref{fig:old_sigma}. The fiducial NSD iso-contour from \citet{sormani22} is shown as the white line representing where the NSD has a density of $6\times10^{7}M_{\odot}/deg^2$. Left panel correspond to the new mean RV map, in comparison with the previous map, the differences are not stunning but expected. Right panel shows the map of the $\sigma_{RV}$, comparing with the previous map, the differences are clear near the galactic plane where new constraints were added. See figures in Appendix \ref{comparison_plots} for a more detailed comparison.}
    \label{fig:new_mean_rv}
\end{figure*}

The final kinematic map of RV, with contours displayed in steps of 
$10\ \mathrm{km}/s$ and the corresponding 
map of the $\sigma_{RV}$, including its contours, is shown in Fig.~\ref{fig:new_mean_rv}. 
Both figures also display the residuals of the fit along Galactic latitude and longitude. 
It is worth saying that the residuals do not exceed 10\% of the model value, 
remaining within the error bars.
Compared with previous kinematic maps in Z14, the updated map of RV exhibits only minor 
differences, particularly at the top edge where more constraints were added. The difference was 
within the expected uncertainties and does not significantly alter the overall cylindrical 
rotation trend. The map of the $\sigma_{RV}$ shows slightly more differences, especially 
near the Galactic  plane. The updated map is better constrained in this region, the absence 
of the “wing-like”  shape along the Galactic plane compared to the previous version in Z14 
(see countours in Fig.~\ref{fig:old_sigma}) is attributed to the additional measurements, 
altering the global shape to a visually boxier/regular morphology. 
The updated map still shows a central $\sigma_{RV}$ peak elongated along the galactic latitude
but smoother than before. A more detailed visual difference with comparison maps could be find in Fig.\ref{fig:diff_meanrv} and Fig.\ref{fig:diff_sigmarv} in the appendix.

We also evaluated the inclusion of external catalogs such as BRAVA and ARGOS, which 
cover regions similar to the outer areas of our survey. 
The comparison revealed differences of $\pm$15 km/s in the mean RV and $\pm$10 km/s 
in $\sigma_{RV}$, with a systematic offset of approximately 4 km/s in $\sigma_{RV}$. 
These discrepancies fall within the expected uncertainties, indicating general consistency
with our velocity map. We decided
not to include them in the present model, given our primary goal of improving constraints 
in the inner regions, which were poorly sampled by these surveys.

\section{Discussion}

From our results, the updated mean RV map confirms the cylindrical rotation of the bulge
reinforced by more the latitude coverage of the new observations. 
This result agrees well with bulge spectroscopical 
surveys such as BRAVA \citep{kunder+2012}, ARGOS \citep{ARGOSsurvey} and Gaia ESO survey
\citep{rojas-arriagada+2014} where the bulge was consistently found to rotate cylindrically. 

The updated $\sigma_{RV}$ map also changed with the new constraints. The $\sigma_{RV}$ peak 
is now smoother than in the previous version, but it is still vertically extended. Near the plane, 
it no longer shows the wings that were observed in the $\sigma_{RV}$ map by Z14. 
Comparisons of the maps by Z14 against models have been made in \citet{gonzalez+16}.
In that case, the model was a pure disk galaxy, simulated by means of an N-body code based on \citet{cole+14}. 
Both the Z14 and the present $\sigma_{RV}$ map agree with that model. The latter shows a
vertically extended peak caused by the elongated orbits of stars in the bar and boxy-peanut.
The shape of the peak depend on the angle between the line of sight and the bar major axis
(in this case $20^\circ$). 
That comparison confirms that the mere presence of a bar, with its orbit anisotropy, can
produce a peak in the projected $\sigma_{RV}$. However, this finding is not sufficient to
explain the peak observed in the data, as the latter could also be produced by a larger
contration of mass in the innermost region.  The model presented in \citet{gonzalez+16}
does not show the -wings- close to the plane that were present in the map by Z14 and
have disappeared now, thanks to the additional observational data.

A more recent work from \citet[][submitted to A\&A]{khoperskov+24} shows RV maps of the MW bulge
using an orbit superposition method. The data were taken from APOGEE DR17, and used to 
calculate and rewind the stellar orbits, assuming the same fixed gravitational potentials 
as in \citet{sormani22} and first released in \citet{portail+17barps}. Different orbit
snapshots were superposed in order to reconstruct the RV kinematics maps, 
specifically the $\sigma_{RV}$ 
map is shown in the Fig. 6 in \citet[][submitted to A\&A]{khoperskov+24}. 
The $\sigma_{RV}$ map was consistent with the central peak vertically elongated, 
and no excess of $\sigma_{RV}$ was shown in the plane as in the present work.

The works mentioned above were based on models of pure disk galaxies, and reproduce 
well the observed kinematics of the bulge. Nonetheless, the presence of a spheroidal component
in the bulge is clearly demostrated by several independent observations 
\citep[for example,][and references therein]{zoccali+2017, lim+21}. In order to understand the origin
of $\sigma_{RV}$ peak, we are currently comparing the present data with the CIELO \citep{tissera+25}
suite of cosmological simulation (Acosta-Tripailao et al. {\it in preparation}).

In closing, we note that the present data do not allow us to impose constraints on the
kinematics of the nuclear stellar disk. Indeed, according to the model of \citet{sormani22},
in the MW the NSD would be very small, as shown by the white contour in the central region of 
Fig.~\ref{fig:new_mean_rv}. The present data are too sparse, in
that region, to include the effect of a NSD. In the future, it would be especially interesting
to fill this region, in order to provide more stringent constraints about the presence (or 
absence) of a clear kinematic signature of the presence of a NSD 
\citep[for example, see discussion in ][]{zoccali+24NSD}.

\section{Summary and conclusions}
\label{sec:summary}
In this study, we presented an update to the known kinematic maps of the MW bulge, 
incorporating 28 new measurements of RV and $\sigma_{RV}$ in the line of 
sight based on MUSE and APOGEE data. 
These measurements complement previous studies, 
such as the GIBS survey and V18, by adding spatial coverage closer to the Galactic plane. 
The updated maps provide new insights into the kinematics of the bulge. Specifically, 
we observe that the $\sigma_{RV}$ contours exhibit a boxier morphology compared to 
previous results, with a significant reduction in the "wing-like" extensions along the 
Galactic plane. The previous $\sigma_{RV}$ central peak studied in V18 is still 
visible, however it is smoother. While the former was more elongated along the 
latitude, likely because now the contours near the plane are better constrained. 

These new findings can be further complemented by studying the separated kinematics of the 
different stellar components in the bulge. Measuring the chemical abundances of the same 
stars included in the present sample would allow a separation into metal-rich and 
metal-poor components, whose corresponding RV in previous works have been shown to be 
different in the inner bulge. The RV indicates a faster rotation for the metal-rich 
population and a slower rotation for the metal-poor one \citep{ness+2013kin_argosIV, 
rojas-arriagada+2020, olivares+24}. Interestingly, the $\sigma_{RV}$ near the plane is higher 
for the metal-rich component than for the metal-poor one, as recently shown by 
\citet{boin24}. This result appears counterintuitive, since the metal-rich population 
is thought to trace the bar structure, where a more coherent rotation—and thus a lower 
$\sigma_{RV}$—would be expected compared to the metal-poor component. Studying the $\sigma_{RV}$
in this context can provide important insights. A higher $\sigma_{RV}$ in the 
metal-rich population may indicate a higher stellar mass density \citep{rix24, horta25}, 
or reflect the anisotropy of the stellar orbits in the bar, the high $\sigma_{RV}$ could be tracing the vertex deviation \citep{babusiaux+10, babusiaux+2016, simion+21}. Furthermore, the differences in $\sigma_{RV}$ constraints formation scenarios consequences like the kinematic fractionation \citep{debattista+2017, fragkoudi+2018}, predicted in the disk secular evolution model, which explains that the differences in the current spatial distribution of the stellar populations depend on the different initial kinematics of these components.

As part of this work, we released a new Python-based photometry package 
{\tt PHOTfun}, which integrates DAOPHOT-II into a GUI to easily perform robust 
PSF photometry of any set of science images, and the extension {\tt PHOTcube} 
that extends the capability of the software to work on 
any IFU datacube in order to extract sources spectrum via PSF photometry. This tool enables 
efficient and robust extraction of spectra from high-density regions, ensuring reliable 
measurements. By reanalyzing previously published datasets, we also emphasized the value of 
using updated techniques to achieve consistent results.

\begin{acknowledgements}

This work is funded by ANID, Millenium Science Initiative, ICN12\_009 awarded to the Millennium Institute of Astrophysics  (M.A.S.), by the ANID BASAL Center for Astrophysics and Associated Technologies (CATA) through grant FB210003, and by  FONDECYT Regular grant No. 1230731. C. Q. Z. acknowledges support from the National Agency for Research and Development (ANID), Scholarship Program Doctorado Nacional 2021 – 21211884, ANID, ESO SSDF 13/23 D grant. A. R. A. acknowledges support from DICYT through grant 062319RA. E. V. acknowledges the Excellence Cluster ORIGINS Funded by the Deutsche Forschungsgemeinschaft (DFG, German Research Foundation) under Germany’s Excellence Strategy – EXC-2094-390783311. A.M. acknowledges support from the project "LEGO – Reconstructing the building blocks of the Galaxy by chemical tagging" (P.I. A. Mucciarelli). granted by the Italian MUR through contract PRIN 2022LLP8TK\_001. A. V. N. acknowledges support from the National Agency for Research and Development (ANID), Scholarship Program Doctorado Nacional 2020 – 21201226, ANID.

\end{acknowledgements}

\bibliographystyle{aa}
\bibliography{BulgeBibliography} 
\begin{appendix} 
\onecolumn
\begin{multicols}{2}
\section{PHOTfun GUI preview.}
\label{photfun}

\begin{figure}[h!]
    \centering
    \includegraphics[width=\hsize]{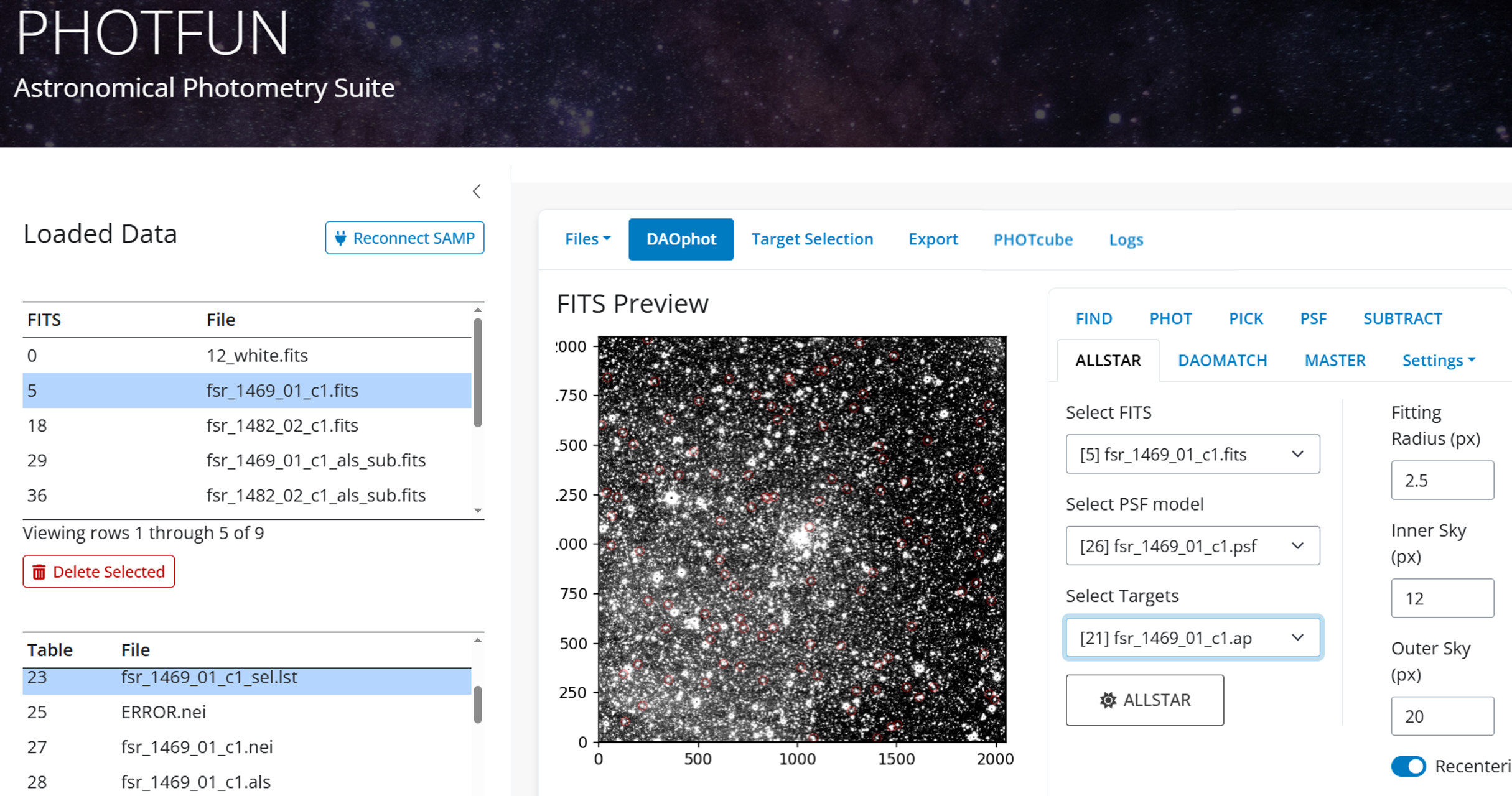}
    \caption{PHOTfun graphical user interface. On the left panel, the loaded files are displayed. At the top center, a set of tabs provides access to the available actions, including the PHOTcube extension. The central bottom panel shows a preview of the selected FITS image, with red circles marking the selected targets. On the right panel, the user can access the available DAOPHOT-II and ALLSTAR routines along with their configurable parameters.}
    \label{fig:photfun_main}
\end{figure}

PHOTfun offers a GUI that streamlines the use of DAOPHOT-II for PSF photometry. As shown in Fig.~\ref{fig:photfun_main}, the interface organizes all core functions into interactive tabs, allowing users to load FITS images, detect sources, model the PSF, and execute photometric analysis using standard DAOPHOT-II routines such as FIND, PHOT, PSF etc. Parameters are adjustable directly within the GUI, and key intermediate results are visualized in real time, thus reducing the reliance on command-line input and making the workflow more accessible and efficient.

\begin{figure}[]
    \centering
    \includegraphics[width=\hsize]{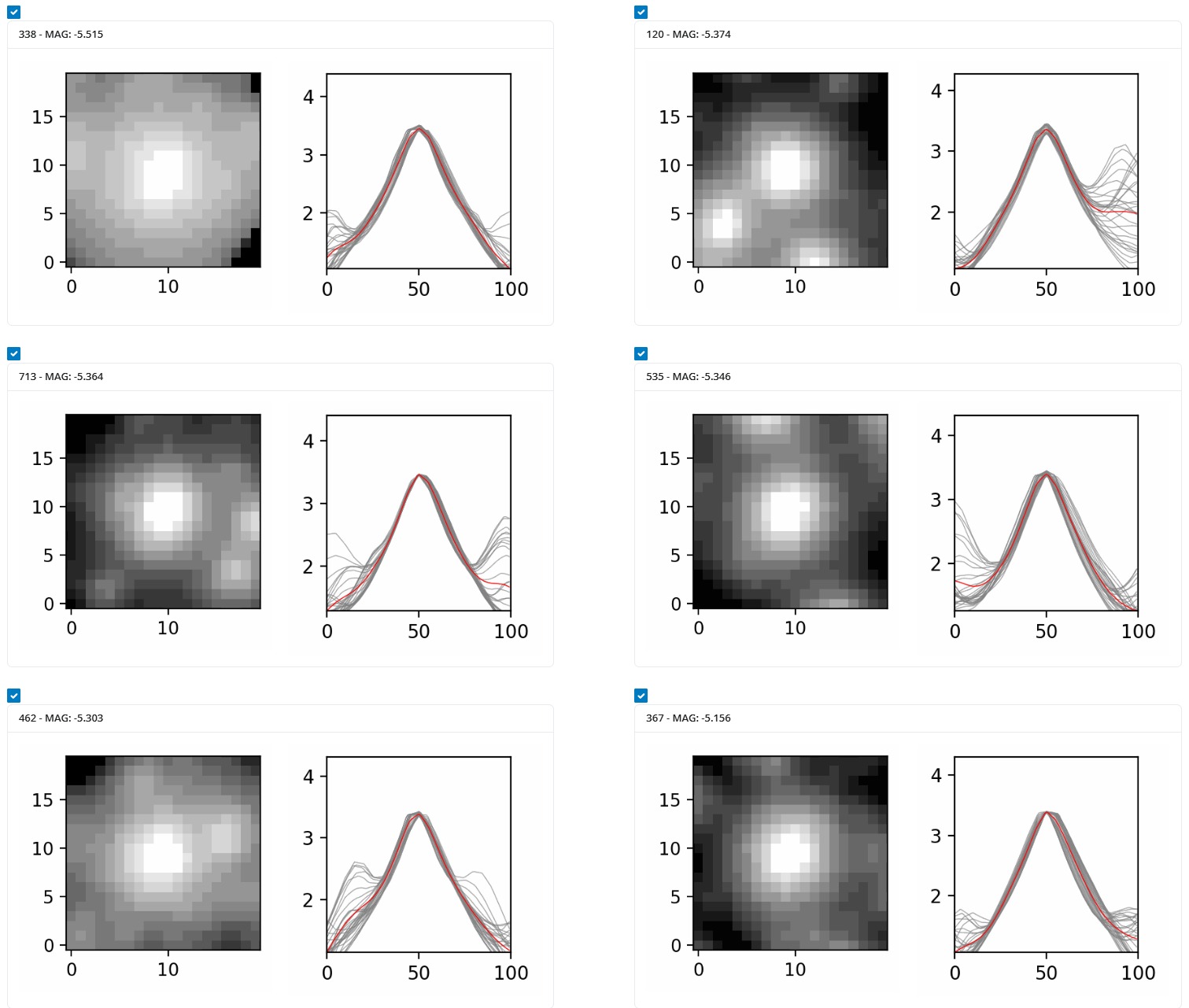}
    \caption{Example of the PSF star selection panel provided by the software. For each candidate star, the interface displays its 20x20 pixels image alongside its light profile, also in 20 pixels (even if the PSF fitting radius is fewer than 10 pixels). Each line in the profile represents a line crossing through the star center, with consecutive lines corresponding to rotated profiles at different angles. The solid line indicates the averaged profile, which ideally approximates a Gaussian shape. Ideally one rejects the profiles which have contamination of near sources in the inner user selected fitting radius.}
    \label{fig:photfun_selection}
\end{figure}

An additional component of PHOTfun is the interactive PSF star selection (Fig.~\ref{fig:photfun_selection}), which displays image cutouts and light profiles for each candidate PSF star. This feature helps users identify and exclude stars with blended or asymmetric profiles. The GUI also supports SAMP\---based communication with external tools like TOPCAT and DS9. 
Additionally, the integrated PHOTcube module extends these capabilities to IFU datacubes, enabling spectra extraction from monochromatic image slices using the same PSF photometric framework.
\end{multicols}

\FloatBarrier
\clearpage

\begin{multicols}{2}
\section{Corner plots for the kinematic maps fitting.}
\label{corner_plots}
\begin{figure*}[!th]
    \centering
    \includegraphics[width=0.8\hsize]{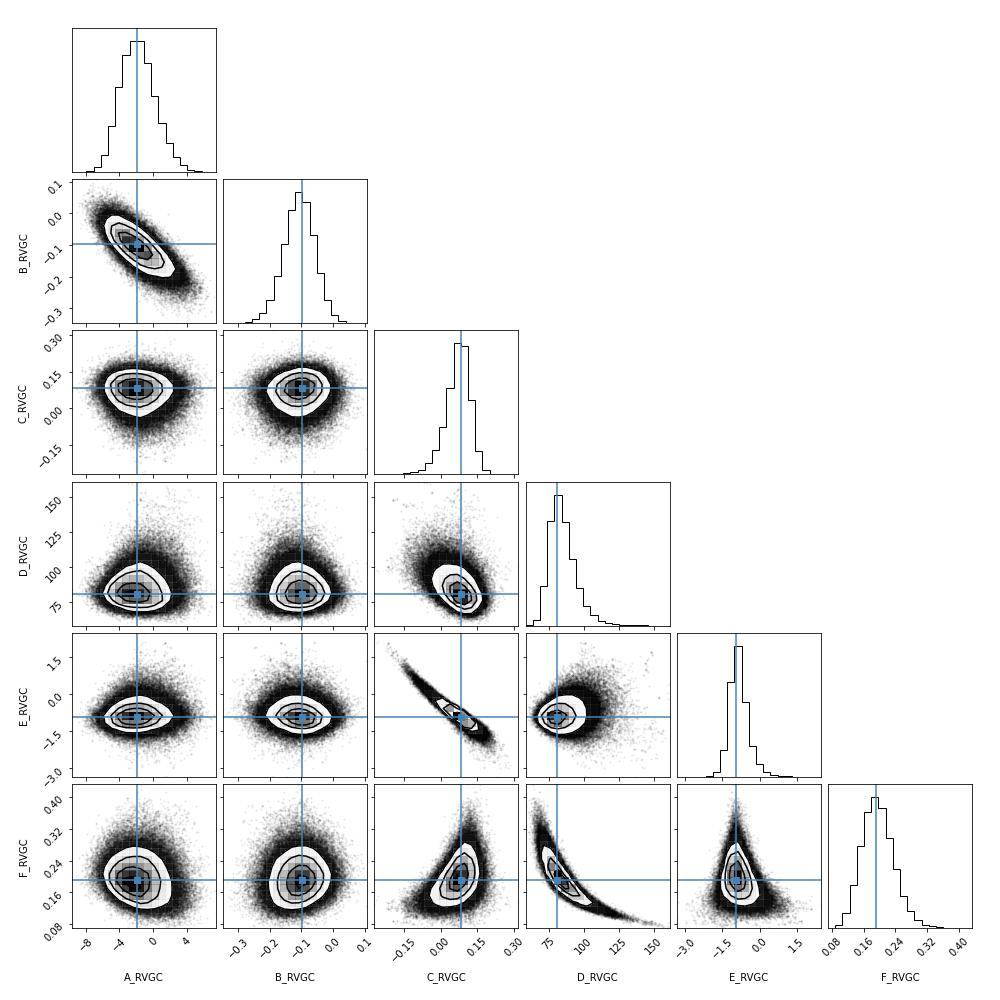}
    \caption{Corner plot showing the posterior distributions of the coefficients for the RV map in the Galactic bulge. These coefficients were derived using MCMC sampling to fit an analytical equation to the updated dataset. The well defined Gaussian shapes in the 2D distributions indicate a robust determination of the coefficients.}
    \label{fig:corner_rvgc}
\end{figure*}

\begin{figure*}[!th]
    \centering
    \includegraphics[width=\hsize]{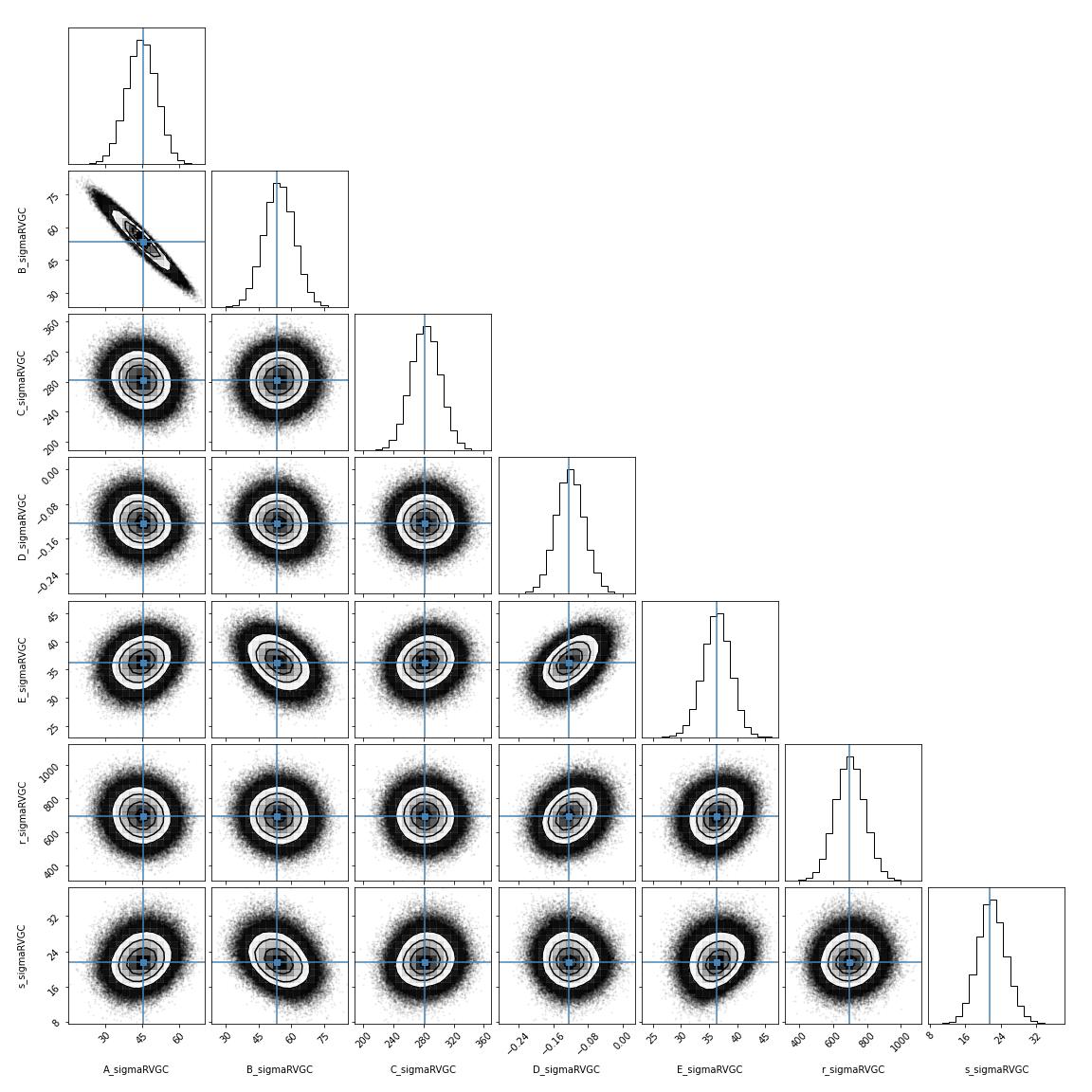}
    \caption{Corner plot showing the posterior distributions of the coefficients for the $\sigma_{RV}$ map in the Galactic bulge. Similar to the previous figure, these distributions represent the robust determination of parameters used to model the $\sigma_{RV}$ profile as we notice, each parameter distribution is well represented as a gaussian.}
    \label{fig:corner_sigmarvgc}
\end{figure*}

To evaluate the robustness of the kinematic model fits, we examined the full posterior distribution of the parameters governing the analytic expressions for the mean radial velocity and velocity dispersion maps. The resulting corner plots from the MCMC sampling are shown in Figures~\ref{fig:corner_rvgc} and \ref{fig:corner_sigmarvgc}. These plots display the parameters distribution along with their covariances, providing a diagnostic tool for identifying degeneracies or parameters that are poorly constrained.

For both the RV and $\sigma_{RV}$ models, the posteriors distribution exhibits well defined Gaussian profiles, indicating that the inferred parameters and their uncertainties are robust. This confirms efficient convergence of the MCMC chains, and validates the statistical reliability of the fitted kinematic profiles. The resulting coefficients, listed in Table~\ref{tab:Coefs}, are thus representative of a stable solution well constrained by the data.
\end{multicols}

\FloatBarrier

\twocolumn
\section{Older vs new maps comparison.}
\label{comparison_plots}
\begin{figure}[!th]
    \centering
    \includegraphics[width=\hsize]{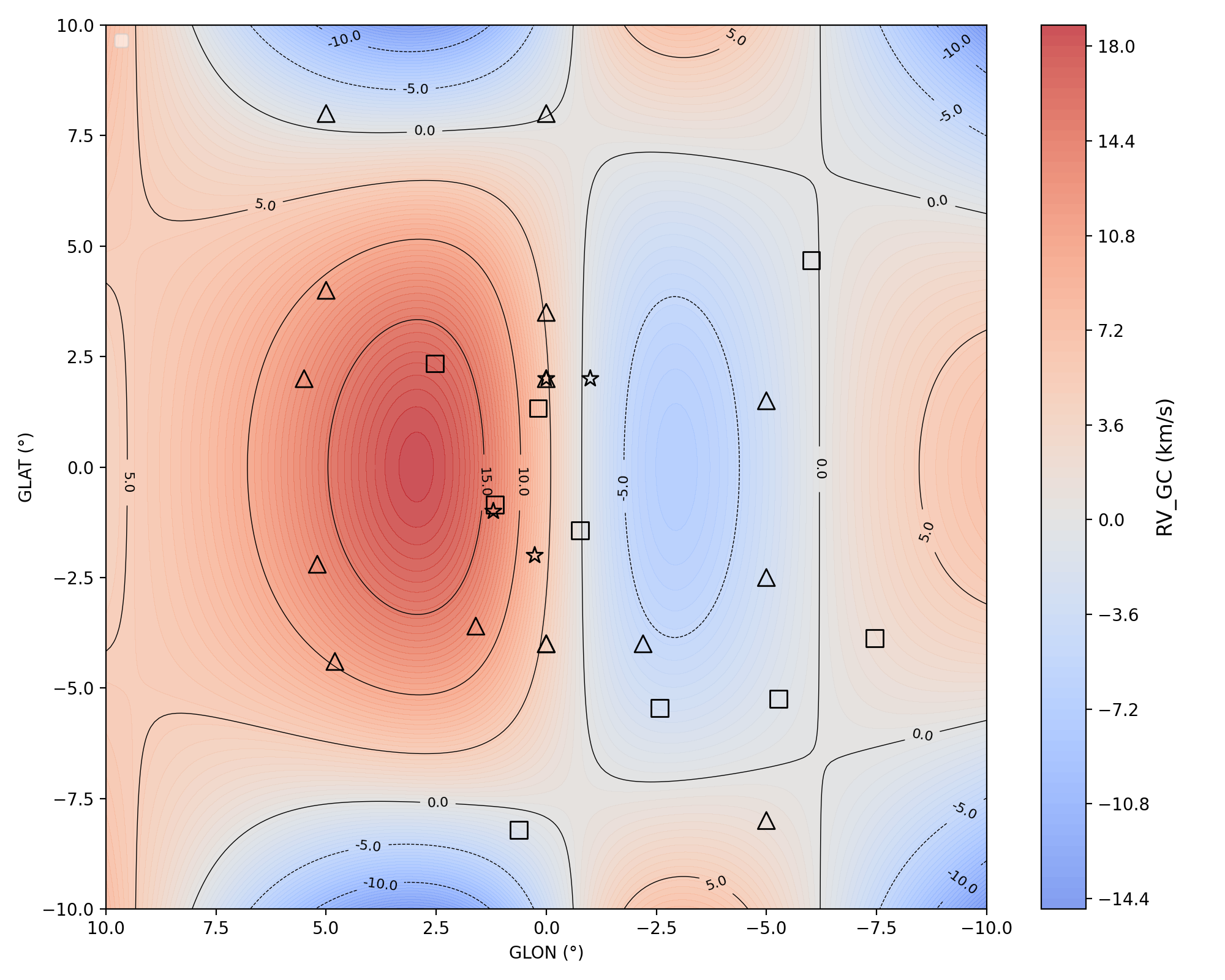}
    \caption{Map of the difference in mean RV between the previous Z14 map and the new one presented here (Z14 - New). The markers follow the same convention as in Fig.~\ref{fig:old_sigma}, but we only show the positions of the new data to highlight the regions where the updated map is better constrained. The new mean RV map is more symmetric and smoother than that of Z14, the asymmetry in the difference map reflects that. The largest differences appear within the inner 2$^\circ$ near the Galactic plane, where most of the new data are located.  }
    \label{fig:diff_meanrv}
\end{figure}

\begin{figure}[!th]
    \centering
    \includegraphics[width=\hsize]{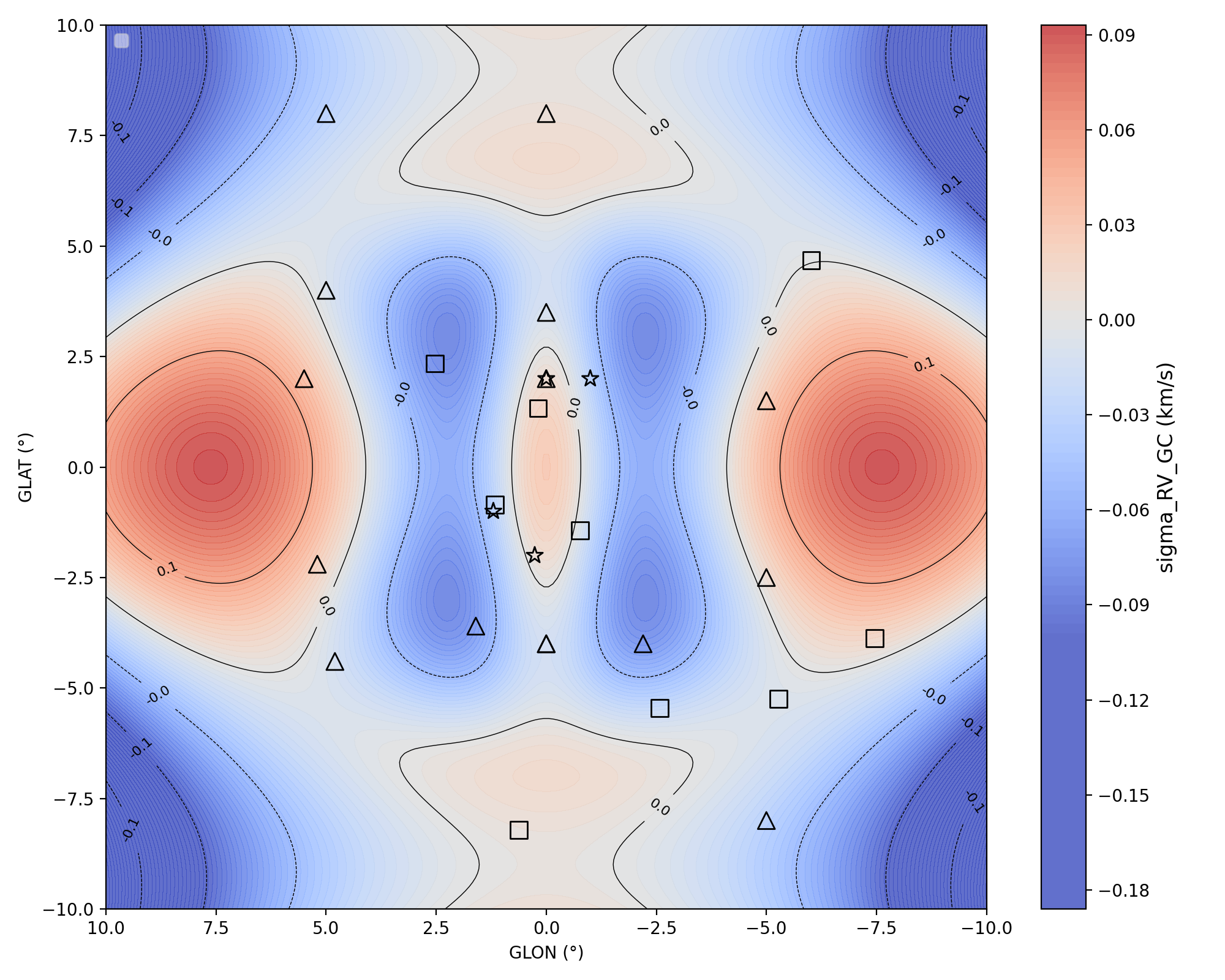}
    \caption{Map of the difference in $\sigma_{RV}$ between the previous Z14 map and the updated version presented here (Z14 - New). The new$\sigma_{RV}$ map shows more significant differences near the Galactic plane, where the additional data have smoothed the central peak and modified the intermediate contours toward a boxier morphology without the "wing-like" shape. }
    \label{fig:diff_sigmarv}
\end{figure}

To highlight the impact of the new MUSE observations and the updated model fits, Figures.~\ref{fig:diff_meanrv} and \ref{fig:diff_sigmarv} present difference maps comparing our results with those of Z14. These maps show the residual in mean RV and velocity dispersion ($\sigma_{RV}$) calculated as Z14 minus the updated values. Only the positions of the new data points (for instance, New MUSE, V18 and APOGEE data) are overlaid, highlighting the regions where our measurements provide improved constraints over the previous dataset.

The comparison shows that most differences are concentrated near the Galactic plane, within $|b| < 2^\circ$, where the new data offer improved spatial coverage. Mean RV differences are generally moderate, mostly under 15~km/s, and correspond to asymmetries in the original Z14 maps that are now corrected. In contrast, the $\sigma_{RV}$ differences are pronounced, highlighting a reshaped morphology of the central velocity dispersion peak. Notably, the updated map removes the wing-like extensions seen in Z14, resulting in a more symmetric and regular structure that better reflects the distribution of the new measurements

\label{comparison_plots}

\end{appendix}

\end{document}